\documentclass{preprint}

\usepackage{subcaption}
\usepackage{tikz}
\usepackage{comment}
\newcommand{\mulimg}[2][]{%
  \begin{tikzpicture}[baseline=(img.south)]
    \node[inner sep=0pt, outer sep=0pt, blend mode=multiply] (img) {\includegraphics[#1]{#2}};
  \end{tikzpicture}}

\lefttitle{Kim et al.}
\righttitle{Unifying vortex-identification criteria and vortex-core extraction}

\title{Unifying vortex-identification criteria and vortex-core extraction}

\author{Seungchan Kim\aff{1}, Tamer A. Zaki\aff{2} \and Donghyun You\aff{1}}

\affiliation{
\aff{1}Department of Mechanical Engineering, Pohang University of Science and Technology (POSTECH), Pohang, Republic of Korea
\aff{2}Department of Mechanical Engineering, Johns Hopkins University, Baltimore, MD, USA
}

\makeatletter
\gdef\@corresau{\textbf{Corresponding authors:} Tamer A. Zaki, \email{t.zaki@jhu.edu}; Donghyun You, \email{dh.you@postech.ac.kr}}
\makeatother

\begin{document}
\maketitle

\begin{abstract}
The differences between vortex cores originally interpreted as pressure-minimum regions and those commonly visualized as regions of high vortex strength are investigated.
To establish a reliable measure of vortex strength, a unified criterion, \(Q_{\mathrm{sp}}\), is introduced by combining the key characteristics of the widely used vortex-identification criteria \(Q\), \(\lambda_2\), and \(\lambda_{ci}\) while addressing their limitations.
Physically, \(Q_{\mathrm{sp}}\) measures the effective attraction or repulsion induced by the square of the velocity-gradient tensor between fluid elements in in-plane vortical motion.
To compare the two vortex-core concepts, two vortex-center extraction methods are constructed according to the respective interpretations.
The pressure-minimum is generalized to a zero-acceleration point at the center of a centripetal-acceleration pattern, whereas the vortex-strength is represented by a local maximum of \(Q_{\mathrm{sp}}\).
The methods are applied to Euler and Stokes flows and to transitional and turbulent channel flows.
Most identified centers form clear centerlines, consistent with filamentary vortex structures, but the two methods yield substantially different results.
The vortex-strength-based method identifies constituent vortices from local concentrations of vortex strength, whereas an acceleration-based center can be interpreted as the center of a resultant vortex arising from the combined influence of surrounding vortices.
The domain of interest required by the acceleration-based method determines the largest characteristic length scale of the vortical motions that influence the resultant vortex.
From a practical perspective, the acceleration-based method, rooted in the original vortex-core concept, is well suited to vortex visualization because it directly considers acceleration and primarily visualizes dynamically significant vortices.

\end{abstract}

\section{Introduction}

\subsection{Vortex-identification methods}
\label{sec:headings}

Although the notion of a vortex appears intuitively straightforward, no universally accepted definition of a vortex has yet been established \citep{panton2024incompressible}.
In this context, numerous vortex-identification criteria have been proposed to identify vortex cores, including the $Q$ criterion \citep{hunt1988eddies}, the $\Delta$ criterion \citep{chong1990general}, and the $\lambda_2$ criterion \citep{jeong1995identification}.
These methods were originally intended to distinguish vortical from non-vortical regions according to the sign of the corresponding criterion.
In subsequent applications, specific nonzero thresholds were imposed on these criteria to identify and visualize strong vortical regions.
For example, the $\Delta$ criterion proposed by \citet{chong1990general} was extended to the $\lambda_{ci}$ criterion, which explicitly represents the local swirling strength \citep{zhou1999mechanisms} and was subsequently applied to visualize vortical structures by \citet{adrian2000vortex} and \citet{adrian2007hairpin}.
Similarly, the $\lambda_2$ criterion was applied to the visualization of vortical structures in a subsequent study by \citet{jeong1997coherent}.
All three criteria, namely $\lambda_{ci}$, $\lambda_2$, and $Q$, have since been widely employed for vortex identification and visualization.

The notion of a \textit{vortex core} is widely used and is closely related to that of a tubular vortex, although the two terms are not synonymous.
The term \textit{vortex core} is commonly used to denote a region of strong vortical motion, with vortex cores often visualized using isosurfaces of vortex strength, whereas a tubular vortex refers to a coherent vortex with a cylindrical or ribbon-like geometry \citep{jimenez1993structure}.
When visualized using vortex strength, regions of strong vortical motion commonly exhibit tubular geometries, particularly in homogeneous isotropic turbulence \citep{jimenez1993structure} and boundary-layer flows.
Consistent with this close association, the term \textit{vortex core} often implies a filamentary geometry \citep{chakraborty2005relationships}.
\citet{she1991structure} suggested that regions of high vorticity exhibit greater coherence because, as the vorticity intensity increases, the strain field becomes increasingly determined by the local vorticity, effectively isolating the local dynamics from the random background field.

Although a vortex core commonly denotes a  strong vortical region, it was originally interpreted as a pressure-minimum region representing the center of vortical motion from the perspective of centripetal force.
This dual interpretation of the \textit{vortex core} may be related to the origins of vortex-identification methods.
Although the $Q$ and $\lambda_2$ criteria originated from attempts to identify pressure-minimum regions, which were referred to as vortex cores in the respective studies \citep{hunt1988eddies, jeong1995identification}, the criteria themselves are related to the second spatial derivatives of pressure.
As a simple example, for a smooth one-dimensional pressure distribution, \(d^2p/dx^2\geq0\) is a necessary condition for a local pressure minimum, but a local maximum of \(d^2p/dx^2\) does not necessarily occur at the pressure minimum.
Rather, the second derivatives of pressure are directly associated with vortex strength: in incompressible flow, their sum, the pressure Laplacian, is related to \(Q\) through \(\nabla^2p=2\rho Q=\rho\left(\lVert\mathsfbi{\Omega}\rVert^2-\lVert\mathsfbi{S}\rVert^2\right)\), where \(\rho\) is the density and \(\mathsfbi{S}\) and \(\mathsfbi{\Omega}\) are the strain-rate and rotation tensors.
More generally, the second derivatives of pressure divided by density have the same dimensions as vorticity squared.
This distinction raises the question of whether vortex-core identification should focus on pressure-minimum regions or strong vortical regions.
To resolve this ambiguity, the vortex structures defined according to these two interpretations must be directly compared.

Such a comparison, in turn, requires a visualization method that clearly resolves the vortex structures associated with the two vortex-core interpretations.
Isosurfaces of vortex-identification criteria are not suitable for this purpose because their geometrical structures vary with the selected threshold and are therefore not uniquely determined \citep{kida1998identification}.
Alternative methods have therefore been developed to represent vortex structures as thin centerlines, commonly referred to as skeletons or worms.
In a three-dimensional flow, a centerline is generally constructed by connecting center points, each defined as an extremum within a selected two-dimensional plane at each grid point.
\citet{kida1998identification} explicitly defined a plane at each grid point using the eigenvectors of the pressure Hessian and constructed vortex centerlines by connecting the pressure minima identified within these planes.
Various centerline extraction methods have also been developed using planes normal to the real eigenvector of the velocity-gradient tensor.
Within such planes, \citet{sujudi1995identification} identified zero-velocity points associated with swirling motion, whereas \citet{wang2022novel} identified maxima of $\lambda_{ci}$ using an iterative method.
Following the latter approach, \citet{yu2026new} identified maxima of another vortex-identification criterion proposed by \citet{liu2018rortex} using an iterative method.
Alternatively, \citet{tanahashi2001appearanceand} selected a plane perpendicular to the vorticity vector and identified a local maximum of $Q$ within that plane.
Separately, \citet{zhu2019vortex} constructed a vortex centerline by identifying two-dimensional local maxima of $Q$ on the coordinate planes of the grid, rather than on planes defined by the local flow, and connecting them through iterative propagation.
Many of these methods identify vortex centerlines using vortex-identification criteria, but the dependence of the results on the selected criterion can undermine confidence in the identified structures.

In the present study, we first formulate a unified vortex-identification criterion encompassing the $Q$, $\lambda_{ci}$, and $\lambda_2$ criteria.
These three methods are not fundamentally distinct and exhibit similar tendencies in most cases.
However, treating these criteria as entirely separate methods may obscure the physical meaning of a vortex and create unnecessary confusion when selecting among them.
We propose a new unified criterion, \(Q_{\mathrm{sp}}\), defined as \(Q\) evaluated within the swirling plane, which provides a clear physical interpretation and addresses the limitations and counterexamples associated with the existing criteria.
Although \(Q_{\mathrm{sp}}>0\) coincides with the compactness condition \(\lambda_{ci}/|\lambda_{cr}|>\delta\) proposed by \citet{chakraborty2005relationships} when \(\delta=1\), the present formulation provides a physical basis for this specific value of \(\delta\) and quantifies vortex strength rather than merely distinguishing vortical from non-vortical regions.

Based on the two different concepts of a vortex core, we develop acceleration-based and vortex-strength-based methods for extracting vortex centers.
The distinction in terminology, specifically the use of the term "vortex center extraction" stresses that we identify the vortex center points.
The vortex-strength-based method quantifies vortex strength using the newly proposed criterion \(Q_{\mathrm{sp}}\).
The acceleration-based method is developed by extending the concept of a vortex core defined as a pressure-minimum region.
This extension is introduced because fluid motion is governed by the total acceleration rather than solely by the conservative force associated with pressure and because the pressure-based concept is not applicable to Stokes flow.
The two extraction methods are assessed using various examples, including Euler flow, Stokes flow, a transitional channel flow, and a turbulent channel flow, and generally reveal substantially different vortex structures.
The results also show that most identified vortex centers form centerlines, consistent with filamentary vortex structures closely related to tubular vortices.

The remainder of this paper is organized as follows.
Section~2 introduces the $\lambda_{ci}$, $Q$, and $\lambda_2$ vortex-identification criteria based on the velocity-gradient tensor.
Section~3 discusses the limitations of each criterion and presents their natural induction into the unified criterion $Q_{\mathrm{sp}}$.
Section~4 explains the acceleration-based and vortex-strength-based vortex-center extraction methods.
Using these methods, Section~5 visualizes vortex centers in various example flows and discusses the resulting vortex structures.
Finally, the conclusions are presented in Section~6.

\section{Relevant vortex identification criteria}
Two generally accepted minimum requirements for a useful vortex identification criterion are that it (1) is Galilean invariant and (2) excludes simple shear from being identified as a vortex despite its nonzero vorticity.
In this respect, neither the vorticity magnitude alone nor the geometry of pathlines or streamlines is sufficient for vortex identification.
The criteria satisfying these requirements, $\lambda_{ci}$, $Q$, and $\lambda_2$ are among the most widely used.
We classify vortex-identification methods into two categories: velocity-based methods, such as $\lambda_{ci}$, which employ the velocity-gradient tensor, and acceleration-based methods, such as $Q$ and $\lambda_2$, which employ the spatially differentiated momentum equation.

\subsection{Vortex identification from the velocity-gradient perspective: The $\lambda_{ci}$ criterion}

\begin{figure}
\centering

\begin{tikzpicture}[
    x={(1cm,0cm)}, y={(0cm,0.35cm)}, z={(0cm,1cm)},
    >=stealth, line cap=round, line join=round
]
    \def\thetaRci{10}
    \def\thetaRcr{-40}
    \def\thetaLci{-170}
    \def\thetaLcr{-110}

    \pgfmathsetmacro{\nx}{-1.2/sqrt(1.2^2+0.45^2+1.95^2)}
    \pgfmathsetmacro{\ny}{-0.45/sqrt(1.2^2+0.45^2+1.95^2)}
    \pgfmathsetmacro{\nz}{1.95/sqrt(1.2^2+0.45^2+1.95^2)}
    \pgfmathsetmacro{\ux}{1.95/sqrt(1.95^2+1.2^2)}
    \pgfmathsetmacro{\uz}{1.2/sqrt(1.95^2+1.2^2)}
    \pgfmathsetmacro{\vx}{\ny*\uz}
    \pgfmathsetmacro{\vy}{\nz*\ux-\nx*\uz}
    \pgfmathsetmacro{\vz}{-\ny*\ux}

    \pgfmathsetmacro{\rcix}{cos(\thetaRci)}
    \pgfmathsetmacro{\rciy}{sin(\thetaRci)}
    \pgfmathsetmacro{\rcrx}{cos(\thetaRcr)}
    \pgfmathsetmacro{\rcry}{sin(\thetaRcr)}
    \pgfmathsetmacro{\lcix}{\ux*cos(\thetaLci)+\vx*sin(\thetaLci)}
    \pgfmathsetmacro{\lciy}{\vy*sin(\thetaLci)}
    \pgfmathsetmacro{\lciz}{\uz*cos(\thetaLci)+\vz*sin(\thetaLci)}
    \pgfmathsetmacro{\lcrx}{\ux*cos(\thetaLcr)+\vx*sin(\thetaLcr)}
    \pgfmathsetmacro{\lcry}{\vy*sin(\thetaLcr)}
    \pgfmathsetmacro{\lcrz}{\uz*cos(\thetaLcr)+\vz*sin(\thetaLcr)}

    \fill[gray!12]
        plot[domain=0:360, samples=100]
        ({2*cos(\x)},{2*sin(\x)},0) -- cycle;
    \draw[black]
        plot[domain=0:360, samples=100]
        ({2*cos(\x)},{2*sin(\x)},0);

    \fill[gray!35, opacity=0.45]
        plot[domain=0:360, samples=100]
        ({1.65*(\ux*cos(\x)+\vx*sin(\x))},
         {1.65*\vy*sin(\x)},
         {1.65*(\uz*cos(\x)+\vz*sin(\x))}) -- cycle;
    \draw[black, dashed]
        plot[domain=0:360, samples=100]
        ({1.65*(\ux*cos(\x)+\vx*sin(\x))},
         {1.65*\vy*sin(\x)},
         {1.65*(\uz*cos(\x)+\vz*sin(\x))});

    \draw[->, thick, gray] (0,0,0) -- (0,0,2.2)
        node[above] {$\boldsymbol{v}_{r}^{L}$};
    \draw[->, thick, black] (0,0,0) -- (-1.2,-0.45,1.95)
        node[above left] {$\boldsymbol{v}_{r}^{R}$};

    \draw[->, thick] (0,0,0) -- ({2*\rcix},{2*\rciy},0)
        node[right] {$\boldsymbol{v}_{ci}^{R}$};
    \draw[->, thick] (0,0,0) -- ({2*\rcrx},{2*\rcry},0)
        node[below right] {$\boldsymbol{v}_{cr}^{R}$};

    \draw[->, thick, gray] (0,0,0) --
        ({1.65*\lcix},{1.65*\lciy},{1.65*\lciz})
        node[below left] {$\boldsymbol{v}_{ci}^{L}$};
    \draw[->, thick, gray] (0,0,0) --
        ({1.65*\lcrx},{1.65*\lcry},{1.65*\lcrz})
        node[below right] {$\boldsymbol{v}_{cr}^{L}$};

    \draw[gray]
        ({0.25*\rcix},{0.25*\rciy},0) --
        ({0.25*\rcix},{0.25*\rciy},0.25) --
        (0,0,0.25);

    \draw[black]
        ({0.25*\lcix},{0.25*\lciy},{0.25*\lciz}) --
        ({0.25*(\lcix+\nx)},{0.25*(\lciy+\ny)},
         {0.25*(\lciz+\nz)}) --
        ({0.25*\nx},{0.25*\ny},{0.25*\nz});
\end{tikzpicture}
\caption{Illustration of the nonorthogonality of the eigenvectors of the velocity gradient tensor $\mathsfbi{A}$ when it has a complex-conjugate eigenvalue pair.
The left and right eigenvectors satisfy $(\boldsymbol{v}^{L})^{\mathsf T}\mathsfbi{A}=\lambda(\boldsymbol{v}^{L})^{\mathsf T}$ and $\mathsfbi{A}\boldsymbol{v}^{R}=\lambda\boldsymbol{v}^{R}$, respectively.
Although the corresponding left and right eigenvalues are identical, their eigenvectors generally differ because $\mathsfbi{A}$ is nonsymmetric.
The real and imaginary parts of a complex right eigenvector, $\boldsymbol{v}_{cr}^{R}$ and $\boldsymbol{v}_{ci}^{R}$, span a plane perpendicular to the real left eigenvector $\boldsymbol{v}{r}^{L}$.
Conversely, the real and imaginary parts of a complex left eigenvector, $\boldsymbol{v}_{cr}^{L}$ and $\boldsymbol{v}_{ci}^{L}$, span a plane perpendicular to the real right eigenvector $\boldsymbol{v}_{r}^{R}$.}
\label{fig:vgt_eigenvectors}
\end{figure}
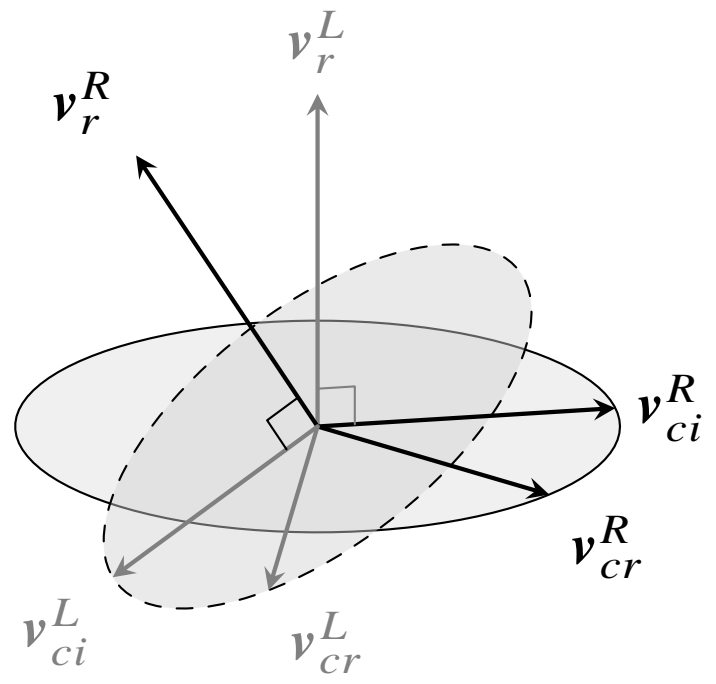

\subsubsection{Non-orthogonality of velocity-gradient tensor}
\label{sec:Non-orthogonality}
The local streamline pattern of the relative velocity around a point can be understood through the velocity-gradient tensor (hereafter denoted by the VGT or \(\mathsfbi{A}\)), whose components are defined as \(A_{ij}=\partial u_i/\partial x_j\), where \(u_i\) is the$i^{\textrm{th}}$ component of the velocity.
The VGT is generally non-symmetric, and its eigenvectors are therefore generally non-orthogonal.
The case in which the VGT has one real right eigenvector and a complex-conjugate pair of right eigenvectors is illustrated in figure~\ref{fig:vgt_eigenvectors}.
The real and imaginary parts of a complex eigenvector span a two-dimensional swirling plane.
The term \textit{swirling plane} reflects two properties: the imaginary part of the complex eigenvalues gives rise to swirling motion \citep{zhou1999mechanisms}, and relative-velocity vectors generated by displacements within this plane also lie in the plane, producing in-plane motion.
This terminology is adopted hereafter.
A relevant property of the left eigenvectors is that the two-dimensional swirling plane is normal to the real left eigenvector, whereas the axis defined by the real right eigenvector is normal to the plane spanned by the real and imaginary parts of a complex left eigenvector.
Owing to this non-orthogonality, a vortex can be understood by focusing either on the two-dimensional swirling plane or on the axis defined by the real right eigenvector.
The interpretation of $\lambda_{ci}$ is based on the two-dimensional swirling plane, whereas another vortex-identification criterion, Liutex \citep{liu2018rortex}, focuses on the axis defined by the real right eigenvector.
In this section, only $\lambda_{ci}$ is discussed, while a detailed explanation of Liutex is provided in Appendix~\ref{app:liutex}.
Hereafter, ``eigenvector'' without a qualifier refers to a right eigenvector.

\subsubsection{The $\lambda_{ci}$ criterion} 
\label{lambdaci}
The quantity $\lambda_{ci}$, which is the imaginary part of the complex eigenvalues of the VGT, quantifies the local swirling strength.
The three-dimensional VGT has either three real eigenvalues or one real eigenvalue and a complex-conjugate pair.
When the VGT has one real eigenvalue and a complex-conjugate pair, \(\lambda_{\pm}=\lambda_{cr}\pm i\lambda_{ci}\), swirling motion occurs within the swirling plane, as discussed in Section~\ref{sec:Non-orthogonality}.

By choosing a coordinate system whose third basis vector is aligned with the real left eigenvector of the VGT, the VGT can be written as
\begin{equation}
\mathsfbi{A}
=
\left[A_{ij}\right]
=
\left[\frac{\partial u_i}{\partial x_j}\right]
=
\begin{pmatrix}
a & b & e \\
c & d & f \\
0 & 0 & g
\end{pmatrix},
\label{eq:VGT_left}
\end{equation}
where the first two basis vectors span the two-dimensional swirling plane, \(a\), \(b\), \(c\), and \(d\) denote the components of the in-plane VGT, and \(e\), \(f\), and \(g\) denote the remaining non-zero components.
The following quantities are defined in terms of the in-plane VGT components:
\begin{equation}
\alpha=\frac{a+d}{2},
\qquad
\beta=\frac{b-c}{2},
\qquad
\gamma=\frac{1}{2}\sqrt{(a-d)^2+(b+c)^2}.
\label{eq:alphabetagamma}
\end{equation}
Here, $\beta$ represents one-half of the vorticity component normal to the swirling plane, while $\alpha$ and $\gamma$ represent the mean and one-half of the difference, respectively, between the two principal strain rates within the swirling plane.
To provide a more intuitive interpretation of these variables, an appropriate rotation can be applied within the swirling plane, after which the VGT takes the following form:
\begin{equation}
\begin{pmatrix}
\alpha+\gamma & \beta & e' \\
-\beta & \alpha-\gamma & f' \\
0 & 0 & g
\end{pmatrix}.
\label{eq:VGT_left2}
\end{equation}
The definitions in equation~\eqref{eq:alphabetagamma} are used throughout the remainder of this paper.
The three eigenvalues of the VGT are then given by
\begin{equation}
\lambda_r=g,
\qquad
\lambda_{\pm}
=
\alpha\pm\sqrt{\gamma^2-\beta^2}.
\end{equation}
For an incompressible flow, $2\alpha+g=0$, and hence $\lambda_r=-2\alpha$.
When $\beta^2>\gamma^2$, the two in-plane eigenvalues form a complex-conjugate pair:
\begin{equation}
\lambda_{\pm} = \lambda_{cr}\pm\lambda_{ci}
= \alpha\pm i\sqrt{\beta^2-\gamma^2}.
\end{equation}
Accordingly, the imaginary part is given by
\begin{equation}
\lambda_{ci} = \sqrt{\beta^2-\gamma^2}.
\label{eq:lci}
\end{equation}
If the VGT has no complex-conjugate pair of eigenvalues, \(\lambda_{ci}\) is defined as zero.

Following the notation by \citet{zhou1999mechanisms}, let the eigenvectors corresponding to the complex eigenvalues be expressed as $\boldsymbol{v}_{cr}\pm i\boldsymbol{v}_{ci}$.
The real and imaginary parts of the complex eigenvector, $\boldsymbol{v}_{cr}$ and $\boldsymbol{v}_{ci}$, are used as non-orthogonal basis vectors for the local coordinates $(y_2,y_3)$ in the swirling plane.
In these coordinates, \citet{zhou1999mechanisms} expressed the instantaneous local streamlines as

\begin{align}
y_2(t)&=\exp(\lambda_{cr}t)[C_c^{(1)}\cos(\lambda_{ci}t)+C_c^{(2)}\sin(\lambda_{ci}t)], \label{eq:lci_zhou_y2}\\
y_3(t)&=\exp(\lambda_{cr}t)[C_c^{(2)}\cos(\lambda_{ci}t)-C_c^{(1)}\sin(\lambda_{ci}t)]. 
\label{eq:lci_zhou}
\end{align}

Here, $C_c^{(1)}$ and $C_c^{(2)}$ are constants determined by the initial position.
With a frozen-VGT assumption, under which the VGT remains unchanged, $\lambda_{ci}$ represents the time-averaged angular frequency of the in-plane swirling motion, with a period of $2\pi/\lambda_{ci}$, thereby motivating its use as a measure of local swirling strength.

\subsection{Vortex identification from the acceleration perspective: The $Q$ and $\lambda_2$ criteria}
The $Q$ and $\lambda_2$ criteria can be classified as methods for understanding vortex cores from an acceleration perspective, particularly in relation to centripetal acceleration.
The analysis begins with the VGT evolution equation \citep{vieillefosse1982local,cantwell1992exact}, 
\begin{equation}
a_{i,j}=\frac{D A_{ij}}{Dt}+A_{ik}A_{kj},
\label{eq:vgt_evolution}
\end{equation}
where \(a_{i,j}\) denotes a component of the acceleration-gradient tensor (AGT).
This equation can be derived by spatially differentiating the momentum equation; the same operation also appears in the derivation of $\lambda_2$ and of the relation $\nabla^2p=2\rho Q$ for incompressible flow.
Both \(Q\) and \(\lambda_2\) focus on the square of the VGT, \(A_{ik}A_{kj}\), through different operations applied to the VGT evolution equation.
Although \(A_{ik}A_{kj}\) is the negative of the self-stretching term, it is interpreted here as the contribution to the acceleration-gradient tensor required for the VGT to remain materially constant.
In other words, the deviation of the AGT from \(\mathsfbi{A}^{2}\) determines the material variation of the VGT.

Taking the trace of the VGT evolution equation yields the following relation associated with the $Q$ criterion:
\begin{equation}
a_{i,i}
=
\frac{D A_{ii}}{D t}
+ A_{ik}A_{ki}.
\label{eq:Q_vgt1}
\end{equation}
The $Q$ criterion is defined as $Q=-\frac{1}{2}A_{ik}A_{ki}$, where $A_{ik}A_{ki}$ is the contribution to the trace of the AGT that remains after the material-derivative term $D A_{ii}/Dt$ is removed.
In other words, the deviation of the AGT trace from $A_{ik}A_{ki}$ determines the material variation of the VGT trace.

Taking the symmetric part of the VGT evolution equation yields the following relation associated with the $\lambda_2$ criterion:
\begin{equation}
\frac{1}{2}\left(a_{i,j}+a_{j,i}\right)
=
\frac{D S_{ij}}{D t}
+
\left(\mathsfbi{S}^2+\mathsfbi{\Omega}^2\right)_{ij}.
\end{equation}
The $\lambda_2$ criterion considers $(\mathsfbi{S}^2+\mathsfbi{\Omega}^2)_{ij}$, which is the contribution to the symmetric part of the AGT that remains after the material variation of the strain-rate tensor, $D\mathsfbi{S}/Dt$, is removed.

\subsubsection{The $Q$ criterion}
In the original study by \citet{hunt1988eddies}, the $Q$ criterion is defined as
\begin{equation}
Q
\equiv
-\frac{1}{2}u_{i,j}u_{j,i}
=
\frac{1}{2}
\left(
\lVert\mathsfbi{\Omega}\rVert^2
-
\lVert\mathsfbi{S}\rVert^2
\right),
\label{eq:Qcriterion}
\end{equation}
where $\mathsfbi{S}$ and $\mathsfbi{\Omega}$ denote the symmetric and antisymmetric parts of the VGT, respectively.
The tensor $\mathsfbi{S}$ is the strain-rate tensor, whereas $\mathsfbi{\Omega}$ is the rotation tensor.
Therefore, $Q>0$ identifies a region in which rotation dominates over strain, i.e., $\lVert\mathsfbi{\Omega}\rVert>\lVert\mathsfbi{S}\rVert$.
Additionally, the second invariant of the VGT can be written as
\begin{equation}
Q_{\mathrm{II}}
\equiv
\frac{1}{2}
\left[
\left(u_{i,i}\right)^2-u_{i,j}u_{j,i}
\right].
\label{eq:second_invariant}
\end{equation}
For an incompressible flow, $u_{i,i}=0$, and hence $Q_{\mathrm{II}}=Q$.
From equation~\eqref{eq:Q_vgt1},
\begin{equation}
a_{i,i}
=
\frac{D A_{ii}}{Dt}
-
2Q.
\label{eq:q_agt}
\end{equation}
For an incompressible flow, $A_{ii}=0$, and the divergence of the momentum equation gives $\rho a_{i,i}=-\nabla^2p$.
Therefore,
\begin{equation}
\nabla^2p
=
2\rho Q.
\end{equation}
Because of this relation between $Q$ and the second derivatives of pressure, $Q>0$ provides only a necessary, but not sufficient, condition for a local pressure minimum in incompressible flow.
For this reason, \citet{hunt1988eddies} have employed an additional condition for identifying vortex cores by explicitly detecting low-pressure regions based on a threshold in their original study.
Without the incompressibility assumption, $Q$ can, in a broad sense, be interpreted in terms of the AGT trace after the contribution from the material variation of the VGT trace is removed, as shown in equation~\eqref{eq:q_agt}.

A relationship can be established between $\lambda_{ci}$ and $Q$.
By choosing a coordinate system that gives the VGT the form presented in equation~\eqref{eq:VGT_left2}, $Q$ can be readily expressed as
\begin{equation}
Q
=
\beta^2-\gamma^2-\alpha^2-\frac{1}{2}g^2
=
\lambda_{ci}^2-\lambda_{cr}^2-\frac{1}{2}g^2.
\label{eq:q_alphabetagamma}
\end{equation}
For an incompressible flow, $g=-2\alpha$, and this expression reduces to
\begin{equation}
Q
=
\beta^2-\gamma^2-3\alpha^2
=
\lambda_{ci}^2-3\lambda_{cr}^2.
\end{equation}

\subsubsection{The $\lambda_2$ criterion}
\label{sec:lambda2}
Taking the symmetric part of the VGT evolution equation gives
\begin{equation}
\frac{1}{2}\left(a_{i,j}+a_{j,i}\right)
=
-\frac{1}{\rho} p_{,ij}+\nu \nabla^2  S_{ij}
=
\frac{D S_{ij}}{Dt}
+
S_{ik}S_{kj}
+
\Omega_{ik}\Omega_{kj},
\label{eq:symmetric_agt}
\end{equation}
where $\nu$ is the kinematic viscosity.
By incorporating the viscous contribution into the negative pressure Hessian and excluding the contribution associated with the material variation of the strain-rate tensor, \citet{jeong1995identification} obtained the symmetric tensor \(\mathsfbi{S}^2+\mathsfbi{\Omega}^2\), which can be interpreted as a modified negative pressure Hessian.
For brevity, the combined contribution of the negative pressure Hessian and the viscous term, which constitutes the symmetric part of the AGT, is hereafter referred to as the symmetric part of the AGT.
They argued that a vortex core exists when the intermediate eigenvalue of this tensor, denoted by $\lambda_2$, is negative.
The condition $\lambda_2<0$ implies that the tensor has two negative eigenvalues and is therefore negative definite within the plane spanned by the corresponding eigenvectors.
Consequently, after the unsteady contribution is excluded, the negative definiteness of the symmetric part of the AGT in the plane spanned by the eigenvectors corresponding to its two negative eigenvalues indicates that the local relative-acceleration vector has an inward radial component in every direction within that plane.
The term \textit{relative} is used because the AGT represents the spatial derivatives of acceleration.
This inward relative-acceleration component is consistent with the concept of the centripetal force required to maintain a vortex.
Although \citet{jeong1995identification} did not explicitly define a vortex-strength measure, subsequent studies \citep{jeong1997coherent} have used the magnitude of $\lambda_2$ to quantify and visualize intense vortical structures.

In two dimensions, the condition that $\mathsfbi{A}^2\boldsymbol{x}$ has a radially inward component in every direction is equivalent to imposing the same condition on the symmetric part of $\mathsfbi{A}^2$, $\mathsfbi{S}^2+\mathsfbi{\Omega}^2$.
Therefore, the complete \(\mathsfbi{A}^2\) term in the VGT evolution equation can retain the same centripetal-acceleration interpretation as its symmetric part.
Let the VGT be written as $\mathsfbi{A}=\left(\begin{smallmatrix}a&b \\ c&d\end{smallmatrix}\right)$, with $\alpha$, $\beta$, and $\gamma$ defined as in equation~\eqref{eq:alphabetagamma}.
This condition is satisfied when $\mathsfbi{S}^2+\mathsfbi{\Omega}^2$ is negative definite, which gives
\begin{equation}
\beta^2-(\gamma+|\alpha|)^2 >0.
\label{eq:centripetal_condition}
\end{equation}

In three dimensions, however, the condition of a radially inward component of $\mathsfbi{A}^2\boldsymbol{x}$ in every direction within a specific plane is not equivalent to the corresponding condition for the symmetric part of $\mathsfbi{A}^2$, $\mathsfbi{S}^2+\mathsfbi{\Omega}^2$, because the specific planes considered for the two tensors are no longer identical.
For $\mathsfbi{S}^2+\mathsfbi{\Omega}^2$, the plane associated with centripetal acceleration is spanned by the eigenvectors corresponding to its two negative eigenvalues.
In contrast, because $\mathsfbi{A}^2$ is non-symmetric, defining an analogous plane is not straightforward.
The appropriate plane for the complete $\mathsfbi{A}^2$ tensor is addressed through the unified vortex-identification criterion introduced in Section~\ref{Unified method}.

\section{Limitations of existing vortex-identification criteria and the proposed unified criterion}
\subsection {The $\lambda_{ci}$ criterion}

The quantity $\lambda_{ci}$ cannot represent the strength of swirling motion when the VGT varies with time, and even a frozen VGT is not physically feasible under certain conditions.
Consider again the instantaneous local streamlines given by \eqref{eq:lci_zhou_y2}--\eqref{eq:lci_zhou}.
As discussed in Section~\ref{lambdaci}, where $\lambda_{ci}$ was introduced, under the frozen-VGT condition, $\lambda_{ci}$ represents the time-averaged angular velocity of the in-plane swirling motion because the time required for one revolution is $2\pi/\lambda_{ci}$.
However, it should be emphasized that only instantaneous data, specifically the VGT at a given instant, are used here.
The only feature that can be inferred from the instantaneous VGT is the local streamline pattern centered at each point, which does not necessarily represent a pathline.
Under the frozen-VGT assumption, the streamline may represent the corresponding pathline; however, the VGT generally varies with time.

Moreover, when the spatial gradient of the viscous force vanishes, as in an Euler flow or a flow with a spatially uniform VGT, maintaining a frozen VGT is physically impossible when \(\lambda_{cr}\neq0\).
Recall the VGT evolution equation, $a_{i,j}=D A_{ij}/Dt+A_{ik}A_{kj}$.
The upper-left $2\times2$ block of $\mathsfbi{A}$ in equation~\ref{eq:VGT_left2} is
\begin{equation}
\mathsfbi{A}=\begin{pmatrix}\alpha+\gamma&\beta\\-\beta&\alpha-\gamma\end{pmatrix}.
\end{equation}
The upper-left $2\times2$ block of $\mathsfbi{A}^2$ is then obtained as
\begin{equation}
\mathsfbi{A}^2=\begin{pmatrix}(\alpha+\gamma)^2-\beta^2&2\alpha\beta\\-2\alpha\beta&(\alpha-\gamma)^2-\beta^2\end{pmatrix}.
\end{equation}
When both the vorticity represented by \(\beta\) and the in-plane expansion or contraction represented by \(\alpha\) are nonzero, \(\mathsfbi{A}^{2}\) contains an antisymmetric component.
For \(D\mathsfbi{A}/Dt\) to vanish, as required under the frozen-VGT assumption, the AGT must contain an antisymmetric component that balances that of \(\mathsfbi{A}^{2}\).
Such an antisymmetric component can be provided only by the non-symmetric viscous contribution to the AGT.
However, this viscous contribution vanishes in an Euler flow or in a flow with a spatially uniform VGT.

For this reason, when only instantaneous VGT data are available, the only reasonable quantity is the angular average of the instantaneous angular velocity rather than the time-averaged angular velocity represented by \(\lambda_{ci}\).
As derived in Appendix~\ref{app:average_angular_vorticity}, the angular average of the angular velocity is simply \(-\beta\), which differs from \(\lambda_{ci}=\sqrt{\beta^2-\gamma^2}\).
Nevertheless, the merit of the \(\lambda_{ci}\) criterion lies in its identification of an appropriate swirling plane within which two-dimensional swirling motion exists in a three-dimensional flow.

\subsection {The $\lambda_2$ criterion}
There exists an extreme case in which the $\lambda_2$ criterion identifies inward relative acceleration in every direction within a specific plane, although no swirling motion exists.
This limitation was demonstrated by \citet{chakraborty2005relationships} in figure~4 of their paper using the following VGT:
\begin{equation}
\nabla\boldsymbol{u}=\begin{bmatrix}1&-a\sin\theta\sin\phi&a\sin\theta\cos\phi\\a\sin\theta\sin\phi&-\xi/2&-a\cos\theta\\-a\sin\theta\cos\phi&a\cos\theta&\xi/2-1\end{bmatrix}.
\end{equation}

Here, \(a\), \(\theta\), \(\phi\), and \(\xi\) are the parameters from \citet{chakraborty2005relationships} and are used only in this paragraph.
For specific values of $\theta$ and $|\mathsfbi{\Omega}|/|\mathsfbi{S}|$, which is determined by \(a\) and \(\xi\), with $\phi = 0$, $\lambda_{ci}$ is zero, indicating the absence of swirling motion in any plane, whereas $\lambda_2$ is negative, indicating centripetal acceleration within a specific plane.
To independently verify this result, the case $\theta/\pi=0.493$, $\xi=0.1$, and $|\mathsfbi{\Omega}|/|\mathsfbi{S}|=0.982$ was examined.
We further examined whether swirling motion could be identified within the specific plane determined by the $\lambda_2$ criterion.
The specific plane is spanned by the eigenvectors corresponding to the two negative eigenvalues, \(\lambda_2\) and \(\lambda_3\), of \(\mathsfbi{S}^2+\mathsfbi{\Omega}^2\), where \(\lambda_1\), \(\lambda_2\), and \(\lambda_3\) denote its eigenvalues in descending order.
This examination was performed despite the absence of in-plane two-dimensional swirling motion in the original VGT.
For this examination, the coordinate system was defined using the eigenvectors corresponding to $\lambda_1$, $\lambda_2$, and $\lambda_3$, respectively.
The VGT was then projected onto the specified plane by simply neglecting the VGT components $A_{11}$, $A_{12}$, $A_{13}$, $A_{21}$, and $A_{31}$ and retaining only the following $2\times2$ block:
\begin{equation}
\begin{pmatrix}A_{22}&A_{23}\\A_{32}&A_{33}\end{pmatrix}.
\end{equation}
However, this projected tensor has two real eigenvalues and therefore does not exhibit swirling motion, even though the examination was not limited to strictly in-plane two-dimensional swirling motion.
Additionally, it is unclear whether \(\lambda_2\) alone can serve as a measure of vortex strength.
The magnitude of the average, \(|(\lambda_2+\lambda_3)/2|\), may provide a more appropriate measure of vortex strength.

\subsection {The $Q$ criterion}
\begin{figure}
    \centering
    \includegraphics[page=1, width=0.45\textwidth]{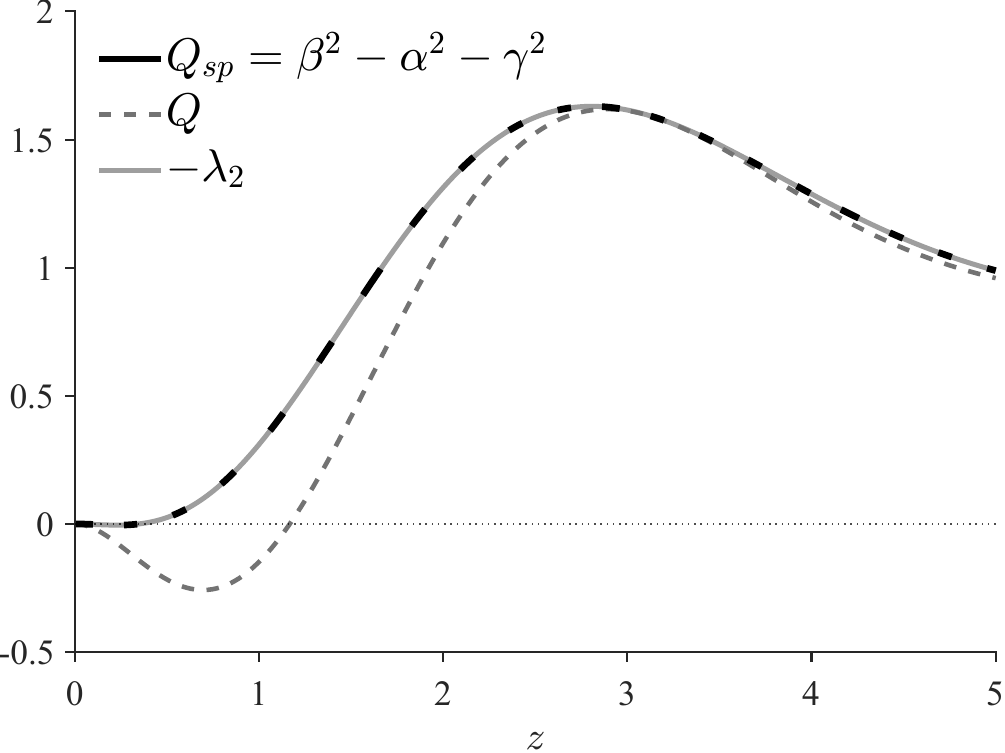}
\caption{Profiles of $Q_{\mathrm{sp}}$, $Q$, and $\lambda_2$ along the $z$ direction for the Bödewadt vortex of \citet{jeong1995identification}; see figure~11(a) of their paper for a schematic of the flow.}
\label{fig:bodewadt_vortex}
\vspace{0.1cm}
\end{figure}

The limitation of the $Q$ criterion was carefully considered by \citet{jeong1995identification}.
For an incompressible flow, the relation \(\nabla^2p=2\rho Q\) indicates whether the average pressure-induced radial acceleration gradient, \(-\boldsymbol{e}_r^{\mathsf T}(\nabla\nabla p)\boldsymbol{e}_r/\rho\), is inward or outward, depending on the sign of \(Q\).
However, this interpretation does not fully correspond to the concept of centripetal force for two reasons: \(Q\) represents only the average of the radial acceleration gradient and therefore does not ensure an inward component in every direction, and it considers all spatial directions rather than only those within a specific plane.
The Bödewadt-vortex example presented by \citet{jeong1995identification}, illustrated schematically in figure~11(a) of their paper, clearly demonstrates this limitation: as shown in figure~\ref{fig:bodewadt_vortex}, \(Q\) is negative at small \(z\) and therefore identifies this region as non-vortical.

This negative value arises from the excessive contribution of vortex stretching or contraction because the second derivative of pressure in the direction normal to the swirling plane is also included.
As shown in equation~\eqref{eq:q_alphabetagamma}, $Q$ can be written as $Q=\beta^2-\gamma^2-\alpha^2-\frac{1}{2}g^2$.
In contrast, the quantity obtained by evaluating \(Q\) in the two-dimensional swirling plane rather than in the full three-dimensional space is hereafter referred to as the swirling-plane \(Q\), denoted by \(Q_{\mathrm{sp}}\), and can be written as
\begin{equation}
Q_{\mathrm{sp}}
=
-\frac{1}{2}u_{l,m}u_{m,l}
=
\beta^2-\gamma^2-\alpha^2 = \lambda_{ci}^2-\lambda_{cr}^2>0,
\label{eq:Qsp}
\end{equation}
where $l,m\in\{1,2\}$ denote the coordinate indices within the swirling plane.
This shows that the term \(-\frac{1}{2}g^2\), which represents vortex stretching or contraction, is included in the three-dimensional \(Q\).
Equivalently, \(Q\) accounts for the strain rate normal to the swirling plane because it contains the magnitude of the three-dimensional strain-rate tensor, as expressed by \(Q = \frac{1}{2}\left(\|\mathsfbi{\Omega}\|^{2}-\|\mathsfbi{S}\|^{2}\right)\).

The Bödewadt-vortex results by \citet{jeong1995identification} are reproduced in figure~\ref{fig:bodewadt_vortex} with \(Q_{\mathrm{sp}}\) included as an additional criterion.
The negative \(Q\) values in the small-\(z\) region of the Bödewadt vortex are eliminated when \(Q_{\mathrm{sp}}\) is used.
\(Q_{\mathrm{sp}}\) will serve as the unified criterion for \(\lambda_{ci}\), \(\lambda_2\), and \(Q\); its physical meaning and a detailed explanation are provided in the following subsection.

\subsection{Unified criterion, $Q_{\mathrm{sp}}$}
\label{Unified method}
The three vortex-identification criteria often yield similar results.  However, all three approaches have respective limitations.  
Therefore, computing vortex-strength statistics using one of these criteria or classifying the flow into rotation- and strain-dominated regions using \(Q\) \citep{chevillard2008modeling,meneveau2011lagrangian} may obscure the physical meaning of the resulting statistics or classification.
By combining the advantages and physically meaningful features of the three quantities, we propose a single unified criterion.

An overview of the unified criterion is presented first.
From the \(\lambda_{ci}\) criterion, the swirling plane is adopted to guarantee swirling motion from the velocity perspective.
The acceleration induced by \(\mathsfbi{A}^2\) within the swirling plane remains two-dimensional and in-plane because \(\mathsfbi{A}^2\) preserves the corresponding eigenvectors of the VGT, \(\mathsfbi{A}\).
Because the \(Q\) and \(\lambda_2\) criteria are both based on \(\mathsfbi{A}^2\), this in-plane acceleration can be analyzed following the concepts underlying these two criteria.
As in the \(\lambda_2\) criterion, the requirement that a centripetal-acceleration pattern exist within the plane is retained.
The principal difference is that the analysis is not restricted to the symmetric part of \(\mathsfbi{A}^2\); instead, the swirling plane is selected as the plane of analysis.
The condition for the existence of a centripetal-acceleration pattern is then given by
\begin{equation}
\beta^2-\gamma^2-\alpha^2 = \lambda_{ci}^2-\lambda_{cr}^2 = Q_{\mathrm{sp}}>0.
\label{eq:Qsp2}
\end{equation}
As shown in equation~\eqref{eq:Qsp}, this condition is equivalent to requiring that the two-dimensional \(Q\) criterion evaluated within the swirling plane, denoted by $Q_{\mathrm{sp}}$, be positive.
Furthermore, $Q_{\mathrm{sp}}$ represents the negative angular average of the radial acceleration gradient, defined as \(\boldsymbol{e}_r^{\mathsf T}\mathsfbi{A}^2\boldsymbol{e}_r\).

\subsubsection{The swirling-plane concept underlying \(\lambda_{ci}\)}
Our starting point is that, under the frozen-VGT condition, \(\lambda_{ci}\) represents the angular velocity averaged over one complete revolution, as discussed previously.
By contrast, the instantaneous angular average of the local angular velocity yields one-half of the vorticity, \(\omega/2\), and therefore cannot provide a vortex-strength measure that excludes simple shear.
Nevertheless, when \(\lambda_{ci}>0\), swirling streamlines exist within the swirling plane; therefore, \(\lambda_{ci}>0\) is the requirement for a vortex from the velocity perspective.
For a two-dimensional flow, using the notation \(\mathsfbi{A}=\left(\begin{smallmatrix}a&b\\c&d\end{smallmatrix}\right)\) and the definitions of \(\alpha\), \(\beta\), and \(\gamma\) in \eqref{eq:alphabetagamma}, the three quantities are expressed as \(\lambda_{ci}^2=\beta^2-\gamma^2\), \(Q=\beta^2-\gamma^2-\alpha^2\), and \(-\lambda_2=\beta^2-(\gamma+|\alpha|)^2\), as shown in \eqref{eq:lci}, \eqref{eq:q_alphabetagamma}, and \eqref{eq:centripetal_condition}, respectively.
Therefore, \(\lambda_{ci}^2\) is greater than or equal to both \(Q\) and \(-\lambda_2\), supporting \(\lambda_{ci}>0\) as the least restrictive of the three conditions.
Although \(\lambda_{ci}>0\) may no longer be the least restrictive condition in a three-dimensional flow, it remains a minimum requirement for a vortex to satisfy criteria from both the velocity and acceleration perspectives; accordingly, the subsequent acceleration-based analysis is performed within the swirling plane.

\subsubsection{Incorporating the centripetal-acceleration concept of the \(\lambda_2\) criterion}

We now turn to vortex identification from the acceleration perspective by considering the VGT evolution equation.
Recall the VGT evolution equation~\eqref{eq:vgt_evolution}, obtained by spatial differentiation of the momentum equation:
\begin{equation*}
a_{i,j} = \frac{D A_{ij}}{D t} + A_{ik}A_{kj}.
\end{equation*}
The term \(\mathsfbi{A}^2\) represents the contribution to the AGT required to maintain a materially constant VGT.
The deviation of the AGT from \(\mathsfbi{A}^2\) determines the material variation of the VGT.
The eigenvectors of \(\mathsfbi{A}\) are also eigenvectors of \(\mathsfbi{A}^2\), whose eigenvalues are the squares of the corresponding eigenvalues of \(\mathsfbi{A}\).
Therefore, \(\mathsfbi{A}^2\) maps vectors within the swirling plane onto vectors within the same plane.
When \(\alpha\neq0\), \(\mathsfbi{A}^2\) possesses a complex-conjugate pair of eigenvalues given by
\begin{equation}
\lambda_{\pm}^{\,2} = \lambda_{cr}^2-\lambda_{ci}^2 \pm 2i\lambda_{cr}\lambda_{ci} = \alpha^2+\gamma^2-\beta^2  \pm  2i\alpha\sqrt{\beta^2-\gamma^2}.
\end{equation}

Following the argument underlying the \(\lambda_2\) criterion, centripetal acceleration should exist within the swirling plane.
Within this plane, the simultaneous presence of swirling motion and centripetal acceleration ensures that the vortex requirements from both the velocity and acceleration perspectives are satisfied.
For a complex-conjugate pair of eigenvalues, the presence of centripetal acceleration within the swirling plane can be assessed in two ways:
\begin{enumerate}
\item The radial component of the relative-acceleration vector \(\mathsfbi{A}^2\boldsymbol{x}\), given by \(\boldsymbol{x}^{\mathsf T}\mathsfbi{A}^2\boldsymbol{x}\), is inward in every direction.
\item The relative-acceleration vector field exhibits a pattern that converges toward the center.
\end{enumerate}
When the restriction of \(\mathsfbi{A}^2\) to the swirling plane has two real eigenvalues, these two conditions coincide.
For a complex-conjugate pair of eigenvalues, however, the first condition is generally more restrictive than the second.
Again, consider a two-dimensional flow with \(\mathsfbi{A}=\left(\begin{smallmatrix}a&b\\c&d\end{smallmatrix}\right)\), where \(\alpha\), \(\beta\), and \(\gamma\) are defined as in \eqref{eq:alphabetagamma}.
The first condition is satisfied when the symmetric part of \(\mathsfbi{A}^2\) is negative definite, which requires \(\left(\gamma+|\alpha|\right)^2-\beta^2<0\) (see \eqref{eq:centripetal_condition}).
By contrast, the second condition is satisfied when the real part of the complex-conjugate eigenvalues of \(\mathsfbi{A}^2\) is negative, which requires $\gamma^2+\alpha^2-\beta^2<0$.
The two conditions differ only when \(\alpha\neq0\), that is, when the swirling motion is accompanied by in-plane expansion or contraction.
The second, more intuitive condition is adopted because, as discussed in the following paragraph, it provides a single quantity that both distinguishes vortical from non-vortical regions and quantifies vortex strength.
In summary, centripetal acceleration is considered to exist when the real part of the complex-conjugate eigenvalues of \(\mathsfbi{A}^2\), $\alpha^2+\gamma^2-\beta^2$, is negative.

Compared with the original \(\lambda_2\) criterion, the present approach has two differences.
First, the symmetric part is not taken in the full three-dimensional space, and the selected plane—the swirling plane—therefore differs from that used in the original \(\lambda_2\) criterion.
Because the swirling plane guarantees the existence of the swirling streamline motion identified by the \(\lambda_{ci}\) criterion, cases in which centripetal acceleration is identified despite the absence of swirling motion can be avoided.
Second, a single parameter, \(\alpha^2+\gamma^2-\beta^2\), determines whether a region is vortical and quantifies its vortex strength.
As discussed previously, \(\lambda_2\) can serve as a criterion for determining whether a region is vortical, but it is not fully appropriate for quantifying the vortex strength because it ignores the magnitude of \(\lambda_3\).
However, when the antisymmetric part of \(\mathsfbi{A}^2\) is retained, even within the two-dimensional swirling plane, only one parameter, \(\alpha^2+\gamma^2-\beta^2\), remains.
Furthermore, \(\alpha^2+\gamma^2-\beta^2\) represents the angular average of the radial acceleration gradient, as derived in Appendix~\ref{app:average_angular_acceleration}.
Therefore, the negative of this parameter, \(\beta^2-\alpha^2-\gamma^2\), can be used both to identify vortical regions and to quantify the vortex strength.

\subsubsection{Physical interpretation and implications of \(Q_{\mathrm{sp}}\)}
The quantity \(\beta^2-\alpha^2-\gamma^2\) obtained in \S\,3.4.2 is identical to \(Q_{\mathrm{sp}}\).
Therefore, in the analysis of \(\mathsfbi{A}^2\), the \(Q\)-based analysis within the swirling plane is consistent with the \(\lambda_2\)-based analysis.
Thus, \(Q_{\mathrm{sp}}\) can be regarded as a unified form of the conventional \(\lambda_{ci}\), \(\lambda_2\), and \(Q\) criteria.
From the velocity perspective, although angular averaging is the only reasonable averaging operation applicable to an instantaneous VGT, applying it to the angular velocity eliminates the contribution of the shear strain rate \(\gamma\) and therefore cannot yield a vortex-strength measure that excludes simple shear.
Alternatively, the minimum angular velocity (Liutex) or the time-averaged angular velocity (\(\lambda_{ci}\)) may be used; however, both quantities fail to account for the in-plane expansion or contraction represented by \(\alpha\) and appear no more appropriate as measures of vortex strength than the angularly averaged value.
From the acceleration perspective, \(Q_{\mathrm{sp}}=\beta^2-\alpha^2-\gamma^2\), which is the negative angular average of the radial acceleration gradient, determines whether a centripetal or centrifugal acceleration pattern exists within the swirling plane while excluding simple shear and accounting for \(\alpha\).

Physically, \(Q_{\mathrm{sp}}\) measures the effective attraction or repulsion between fluid elements within the swirling plane induced by \(\mathsfbi{A}^{2}\), the contribution to the acceleration-gradient tensor required for the VGT to remain materially constant.
For \(\lambda_{cr}\neq0\), or equivalently \(\alpha\neq0\), the condition \(Q_{\mathrm{sp}}>0\) corresponds to the compactness condition \(\lambda_{ci}/|\lambda_{cr}|>\delta\) with \(\delta=1\), following the formulation of \citet{chakraborty2005relationships}.
Thus, \(Q_{\mathrm{sp}}>0\) provides a physically based value for \(\delta\).

A single unified criterion has now been established, thereby providing a more consistent basis for quantifying vortex strength.
Accordingly, this unified criterion is used as a measure of vortex strength to extract vortex-strength-based vortex centers.
For example, in the ABC flow \citep{dombre1986chaotic}, the \(Q_{\mathrm{sp}}\)-based vortex centers reveal a distinct structure that differs considerably from the vortex centers obtained using the \(\lambda_{ci}\) and \(\lambda_2\) criteria.
Additionally, as discussed in detail in the next section, the remaining eigenvector of the VGT, which is real and is not used in defining \(Q_{\mathrm{sp}}\), is used as the vortex axis.

\section{Vortex-center extraction methods}
Two types of vortex centers are considered: those defined by locally high vortex strength and those defined as the centers of centripetal-acceleration patterns.
As mentioned in the Introduction, these two methods are considered because vortex-strength isosurfaces are commonly used in practice to identify vortex cores, whereas the original interpretation of a vortex core is associated with centripetal acceleration in the vicinity of a pressure minimum.
Vortex strength is quantified using the proposed unified criterion \(Q_{\mathrm{sp}}\), whereas the pressure-minimum concept is generalized to the center of a centripetal-acceleration pattern, thereby extending the acceleration-based method to a broader class of flows, including Stokes flows.

In a three-dimensional flow, both methods represent vortices by plotting their centers, which generally align to form connected lines, hereafter referred to as vortex \textit{centerlines}, following \citet{wang2022novel}.
The centers are identified using grid points for which the distance to the corresponding estimated center is smaller than a prescribed threshold, with each estimated center lying within a specified two-dimensional plane.
The direction normal to this plane is hereafter referred to as the \textit{axis}, and the plane itself as the \textit{axis-normal plane}.
For the two methods, the center estimated within the axis-normal plane at each grid point is defined, respectively, as a local maximum of the vortex strength \(Q_{\mathrm{sp}}\) and a zero-acceleration point surrounded by a centripetal-acceleration pattern.
Local maxima of vortex strength are used instead of conventional vortex-strength isosurfaces because the latter depend on an arbitrary threshold and identify regions rather than center points.

\subsection{Selection of the axis and axis-normal plane}
For the vortex-strength-based method using \(Q_{\mathrm{sp}}\), the unique real right eigenvector of the VGT is selected as the axis, thereby defining the axis-normal plane as the plane normal to this eigenvector.
The axis-normal plane differs from the swirling plane, which is normal to the real left eigenvector.
Although \(Q_{\mathrm{sp}}\) is defined in the swirling plane, the real right eigenvector is chosen as the axis because the ridge line of \(Q_{\mathrm{sp}}\) is analytically more closely related to this direction.
This relationship is supported by the following property: when the direction of the real eigenvector is uniform throughout a region, $Q_{\mathrm{sp}}$ does not vary along that direction.
To demonstrate this property, consider a coordinate system whose third basis direction, $z$, is aligned with the real eigenvector.
The following relations are then obtained:
\begin{equation}
\partial_z u_x = \partial_z u_y = 0.
\end{equation}
Because the direction of the real right eigenvector is assumed to be spatially uniform, it follows that
\begin{equation}
\partial_z\left(\partial_{l'}u_{m'}\right)=0,
\qquad
l',m'\in{1,2},
\end{equation}
where \(l'\) and \(m'\) denote the indices of two basis directions spanning the axis-normal plane.
In the swirling plane, $Q_{\mathrm{sp}} = -\frac{1}{2}A_{lm}A_{ml}$, where \(l\) and \(m\) denote the indices of two basis directions spanning the swirling plane.
It can be shown that $Q_{\mathrm{sp}}$ is identical to $-\frac{1}{2}A_{l'm'}A_{m'l'}$, which is evaluated in the axis-normal plane.
This equality follows because the tensors obtained by projecting the three-dimensional VGT onto the swirling plane and the axis-normal plane have the same pair of complex-conjugate eigenvalues, as these planes are spanned by the real and imaginary parts of the right and left complex eigenvectors, respectively.
Therefore, $Q_{\mathrm{sp}}$ does not vary along the direction of the real right eigenvector, suggesting that this eigenvector provides an appropriate axis for detecting the ridge line of $Q_{\mathrm{sp}}$.

Additionally, because the eigenvalues associated with the left and right eigenvectors are identical, \(Q_{\mathrm{sp}}\), the existence of projected swirling motion, and the existence of a centripetal-acceleration pattern associated with \(\mathsfbi{A}^{2}\) are equivalent in the swirling and axis-normal planes, although the motion projected onto the axis-normal plane is not generally confined to that plane.
This equivalence supports the use of the real right eigenvector as the vortex axis.
Further support comes from the fact that the real eigenvector does not enter the definition of \(Q_{\mathrm{sp}}\).
To account fully for the non-orthogonality of the VGT, using the real right eigenvector as the axis also appears reasonable.
The present tests show that choosing either the real left eigenvector or the real right eigenvector as the axis does not produce significant differences in the major vortex structures.
For example, when $Q_{\mathrm{sp}}$ has a cylindrical distribution such as $Q_{\mathrm{sp}}=-x^2-y^2$, with no dependence on $z$, any plane that is not strictly parallel to the $z$-axis yields the same center, which lies on the $z$-axis.
However, in regions far from the actual vortex center, where the distribution of $Q_{\mathrm{sp}}$ can deviate substantially from a cylindrical shape, using the real left eigenvector as the axis may produce spurious noise.
This result may be attributed to the large difference between the directions of the real left and right eigenvectors in these regions.
This spurious noise can be observed in the Euler-flow example presented in the next section.

The acceleration-based method uses the same axis and axis-normal plane as the vortex-strength-based method, and this axis is also an eigenvector of the square of the VGT, \(\mathsfbi{A}^{2}\).
The eigenvectors of the AGT could also be considered as candidates for defining the axis.
However, choosing an eigenvector of the AGT would disconnect the axis from the swirling motion, whereas the aim is to identify a centripetal-acceleration pattern within a plane in which swirling motion occurs.
Instead, the AGT is used to determine whether a centripetal-acceleration pattern is present.
Acceleration is described by the momentum equation itself, rather than by its spatial derivative.
Therefore, the search for a zero-acceleration point cannot be separated into a material-derivative term and a remaining term, as in the VGT evolution equation obtained by taking the spatial derivative of the momentum equation.
Accordingly, the associated centripetal-acceleration pattern is assessed using the full AGT rather than only the \(\mathsfbi{A}^{2}\) contribution that remains after the material-derivative term is removed.
Furthermore, assessing the pattern using $\mathsfbi{A}^{2}$ alone is not informative because $\mathsfbi{A}^{2}$ necessarily exhibits a centripetal-acceleration pattern whenever $Q_{\mathrm{sp}}>0$.

\subsection{Identification of vortex centers in the acceleration-based method}

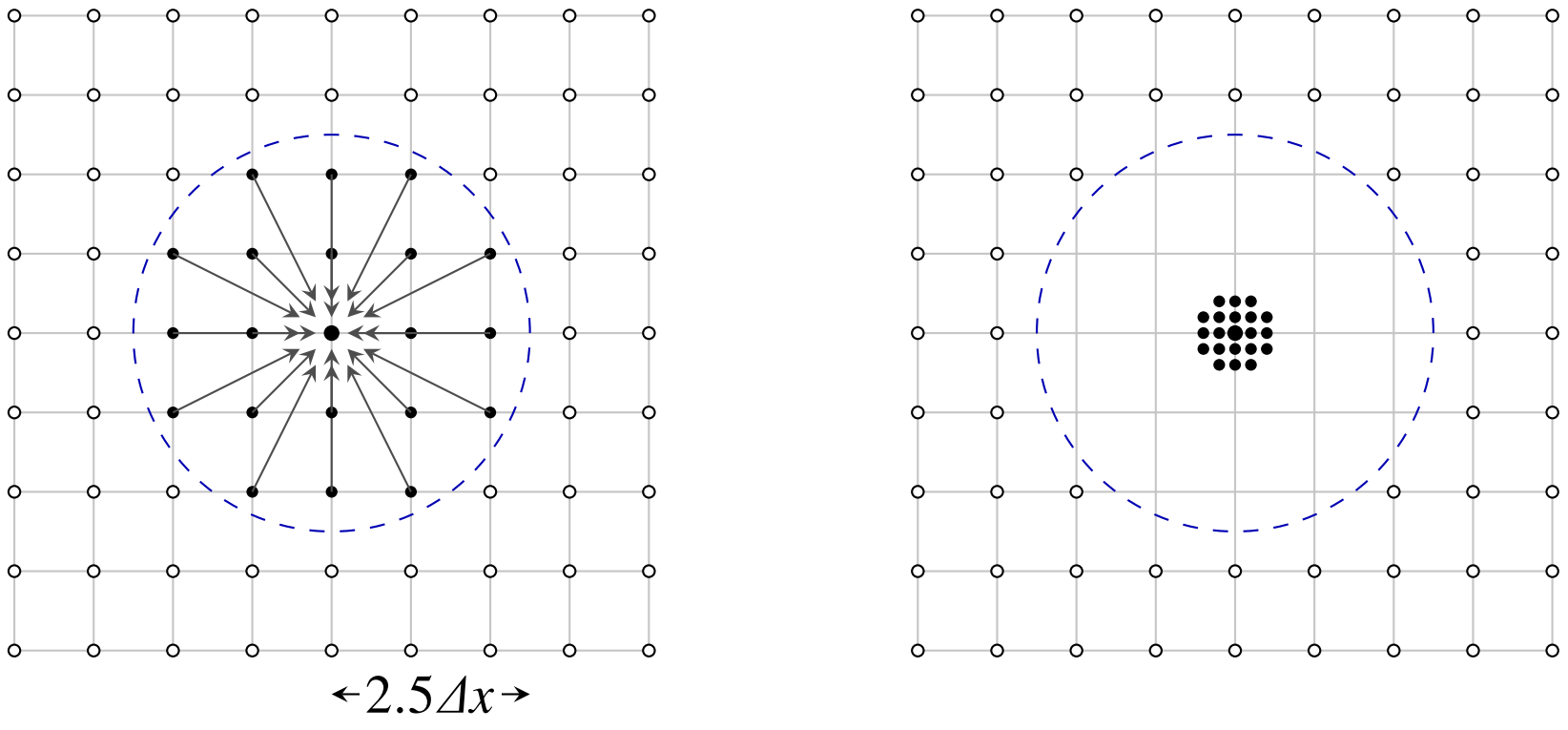
\begin{figure}
\vspace{5pt}
\centering
\def\R{2.5}
\def\moveRatio{0.8}
\def\pointSize{2.1pt}
\def\figureScale{0.55}
\pgfmathsetmacro{\remainRatio}{1-\moveRatio}
\pgfmathsetmacro{\Rsquared}{\R*\R}

\begin{tikzpicture}[scale=\figureScale,>=stealth]
\draw[gray!45,thin] (-4,-4) grid (4,4);
\draw[blue!70!black,dashed] (0,0) circle (\R);
\foreach \i in {-4,...,4}{
    \foreach \j in {-4,...,4}{
        \pgfmathtruncatemacro{\inside}
        {(\i*\i+\j*\j <= \Rsquared) ? 1 : 0}
        \ifnum\inside=1
            \fill[black] (\i,\j) circle (\pointSize);
            \ifnum\i=0
                \ifnum\j=0
                \else
                    \draw[->,white!30!black,thin]
                    (\i,\j) --
                    ({\remainRatio*\i},{\remainRatio*\j});
                \fi
            \else
                \draw[->,white!30!black,thin]
                (\i,\j) --
                ({\remainRatio*\i},{\remainRatio*\j});
            \fi
        \else
            \filldraw[fill=white,draw=black]
            (\i,\j) circle (\pointSize);
        \fi
    }
}
\fill[black] (0,0) circle (2.8pt);
\draw[<->,blue!0!black]
(0,-4.55) -- (\R,-4.55)
node[midway,fill=white,inner sep=1pt] {$\R \Delta x$};
\node[text width=6cm, align=center, font=\footnotesize] at (0,-5.9)
{(a) Points at their grid locations $\boldsymbol{x}$ with estimated radius vector $\boldsymbol{r}$};
\end{tikzpicture}
\hspace{-5pt}
\begin{tikzpicture}[scale=\figureScale,>=stealth]
\draw[gray!45,thin] (-4,-4) grid (4,4);
\draw[blue!70!black,dashed] (0,0) circle (\R);
\foreach \i in {-4,...,4}{
    \foreach \j in {-4,...,4}{
        \pgfmathtruncatemacro{\inside}
        {(\i*\i+\j*\j <= \Rsquared) ? 1 : 0}

        \ifnum\inside=1
            \fill[black]
            ({\remainRatio*\i},{\remainRatio*\j})
            circle (\pointSize);

        \else
            \filldraw[fill=white,draw=black]
            (\i,\j) circle (\pointSize);

        \fi
    }
}
\fill[black] (0,0) circle (2.8pt);
\node[text width=6cm, align=center, font=\footnotesize] at (0,-5.9)
{(b) Points at their estimated center locations $\boldsymbol{x}\!+\!\boldsymbol{r}$ };

\end{tikzpicture}

\caption{
Description of vortex-center visualization.
To visualize the vortex center, only data points for which the magnitude of the estimated radius is less than a prescribed threshold are plotted; the threshold is set to $2.5\Delta x$ in this example.
(a) Data points are plotted at their grid locations, $\boldsymbol{x}$.
The gray arrows represent the estimated radius vectors $\boldsymbol{r}$.
To indicate that the distance cannot be calculated exactly, the magnitude of \(\boldsymbol{r}\) in this schematic is drawn slightly smaller than the actual distance to the vortex center for visual clarity, while its endpoint remains near the  central grid point.
(b) The data points are plotted at their estimated center locations, $\boldsymbol{x}+\boldsymbol{r}$, resulting in a greater concentration of points near the vortex center and making the center more clearly visible.}
\label{fig:grid-displacement}
\end{figure}

Two conditions are required to identify the vortex center corresponding to each grid point.
First, the grid point must lie within a vortical region, i.e. \(Q_{\mathrm{sp}}>0\).
Second, the AGT projected onto the axis-normal plane must produce an inward relative-acceleration pattern.
When the projected AGT has two real eigenvalues, both must be negative.
When it has a complex-conjugate pair of eigenvalues, their real part must be negative.

To visualize the vortex centers, only selected grid points are used.
The grid points are selected if their corresponding estimated vortex centers, defined as the estimated zero-acceleration points within the axis-normal planes, lie within a prescribed distance of the grid points.
This procedure is illustrated in figure~\ref{fig:grid-displacement} and explained in detail below.
The displacement to the zero-acceleration point is estimated using a modified version of the method proposed by \citet{miura1997identification}, which was originally developed to estimate the displacement to a pressure-minimum location within a specified two-dimensional plane.
Let \(\boldsymbol{x}\) denote the current grid point and \(\boldsymbol{x}_0=\boldsymbol{x}+\boldsymbol{r}\) denote the zero-acceleration point, where \(\boldsymbol{r}=(r_x,r_y)\).
When \(|\boldsymbol{r}|\) is sufficiently small, the projected AGT can be assumed to remain constant between \(\boldsymbol{x}\) and $\boldsymbol{x}_0$.
Using a first-order Taylor-series approximation about the current grid point gives
\begin{equation}
\boldsymbol{a}(\boldsymbol{x}_0)  = \boldsymbol{a}(\boldsymbol{x}+\boldsymbol{r}) \approx \boldsymbol{a}(\boldsymbol{x}) + \mathsfbi{B}\boldsymbol{r}
= \boldsymbol{0},
\end{equation}
where \(\boldsymbol{a}\) and \(\mathsfbi{B}\) denote the acceleration and the AGT, respectively, projected onto the axis-normal plane.
The displacement from the current grid point to the zero-acceleration point is therefore
\begin{equation}
\boldsymbol{r} = -\mathsfbi{B}^{-1} \boldsymbol{a}(\boldsymbol{x}).
\end{equation}
Thus, the position of the zero-acceleration point relative to the current grid point can be estimated using a first-order Taylor-series approximation, which requires \(|\boldsymbol{r}|\) to be sufficiently small.
However, this requirement cannot be guaranteed exactly on a discretized grid, particularly in turbulent flow where the flow varies rapidly in space.
Therefore, the terms \textit{estimated displacement to the center} and \textit{estimated center} are more precise.
For simplicity, the former is hereafter referred to as the \textit{estimated radius vector}, and its magnitude as the \textit{estimated radius}.
Note that as the sampled grid point moves farther from the actual zero-acceleration point, this assumption becomes less valid.
The estimated radius is generally larger than the actual distance because the magnitude of the AGT is typically high near the center and gradually decreases outward.

The visualization is constructed by plotting one point for each selected grid point.
To improve the clarity of the vortex-center visualization, each point is plotted at its estimated center position, \(\boldsymbol{x}+\boldsymbol{r}\), as shown in figure~\ref{fig:grid-displacement}(b), rather than at its original grid position, \(\boldsymbol{x}\), as shown in figure~\ref{fig:grid-displacement}(a).

\subsection{Identification of vortex centers in the vortex-strength-based method}
The technique used to identify and visualize these vortex centers is similar to that employed in the acceleration-based method.
First, the grid point must lie within a vortical region, i.e. \(Q_{\mathrm{sp}}>0\); this condition is identical to that in the acceleration-based method.
Second, to identify a local maximum, both eigenvalues of the Hessian of \(Q_{\mathrm{sp}}\) within the axis-normal plane must be negative.
This procedure is simpler than its acceleration-based counterpart because \(Q_{\mathrm{sp}}\) is a scalar field and its Hessian is symmetric, whereas the AGT is generally asymmetric.
The estimated radius used for data selection is calculated following the same procedure as in the acceleration-based method.

One subtlety is that \(Q_{\mathrm{sp}}\) is defined only when the VGT has one real eigenvalue and one complex-conjugate pair, which can make the calculation of its Hessian infeasible in certain situations.
This issue generally does not arise because the condition \(Q_{\mathrm{sp}}=\beta^2-\gamma^2-\alpha^2>0\) is more restrictive than \(\beta^2-\gamma^2>0\), thereby providing a margin between the \(Q_{\mathrm{sp}}>0\) region and the boundary beyond which \(Q_{\mathrm{sp}}\) is undefined.
However, when \(\alpha=0\), these two conditions coincide, leaving no such margin for calculating the Hessian.
Even when \(\alpha\neq0\), \(Q_{\mathrm{sp}}\) may be undefined at a neighboring point required to evaluate its Hessian if the grid spacing in a numerical simulation is too large.
To avoid this computational issue, an additional condition is imposed: a grid point is excluded whenever \(Q_{\mathrm{sp}}\) is undefined at any neighbouring point required for the Hessian calculation.

\subsection{A comparison of acceleration-based and vortex-strength-based vortex centers}
\label{sec:briefcomparison}
A brief conceptual comparison of acceleration-based and vortex-strength-based vortex centers is presented before the two methods are assessed using examples in §~\ref{sec:examples}.
To provide intuition, consider the simplest case: the rigid-body rotation of a cylinder.
In this case, the acceleration-based vortex center coincides with the intuitive center of rotation, the cylinder axis, where the acceleration vanishes and the surrounding centripetal acceleration points toward the axis.
In contrast, rigid-body rotation has uniform vortex strength throughout the domain, so no localized maximum exists from which to identify a vortex-strength-based vortex center.
Similarly, in Stokes flow, the maximum vortex strength does not necessarily occur at the intuitive center of rotation, as demonstrated by the example presented later, and the vortex-strength-based method therefore fails to identify the intuitive center of rotation.

However, the vortex-strength-based method has a clear characteristic: it yields invariant vortex centers for a given flow motion.
Unlike the acceleration-based method, the vortex-strength measure \(Q_{\mathrm{sp}}\) is derived from the spatially differentiated momentum equation and therefore depends on the acceleration gradient rather than on the acceleration itself; it is thus insensitive to spatially uniform absolute acceleration.
In contrast, the acceleration-based method can yield different vortex centers for the same fluid motion, as illustrated by a simple thought experiment.
Consider two isolated boxes, each filled with an incompressible fluid of constant density.
The fluids in the two boxes are subjected to the same initial disturbance.
One box is pulled with a constant acceleration, whereas the other remains stationary.
As time evolves, the internal flow motions in the two boxes should remain identical because, for an incompressible flow, adding a spatially uniform acceleration to the momentum equation does not alter the relative flow dynamics.
In this situation, the vortex-strength-based method yields the same result for both boxes, whereas the acceleration-based method yields different results.

This means that the acceleration-based method cannot yield consistent vortex centers simply by choosing an inertial reference frame.
In principle, an observer can identify a non-inertial reference frame by detecting the acceleration of the frame itself.
However, even when the flows in the pulled and stationary boxes are evaluated in inertial reference frames, the acceleration-based method yields different vortex centers despite their identical internal motions.
As a simpler two-dimensional thought experiment, consider a rigidly rotating disk undergoing uniform translational acceleration.
The disk center remains the intuitive center of rotation, but, in an inertial reference frame, the translational acceleration shifts the zero-acceleration point away from the disk center; if sufficiently large, it can shift the point outside the disk.
For the zero-acceleration point to coincide with the disk center, a reference frame translating with the disk center must be selected.
This selection is analogous to the spatial-mean subtraction employed by \citet{haller2016defining}: there, the mean acceleration over the domain of interest is calculated and subtracted from the acceleration field.
That operation is equivalent to selecting a reference frame that follows the domain of interest rather than an inertial reference frame.
Consequently, an acceleration-based vortex center must be interpreted with respect to a reference frame associated with the designated domain of interest.

From another perspective, when the domain of interest and its associated reference frame are clearly specified, an acceleration-based vortex center may help characterize the global fluid motion because the method deals directly with acceleration.
Once the zero-acceleration points are identified as vortex centers and connected to form a centerline, the acceleration field surrounding the centerline can be approximately mapped.
A detailed comparison of the two methods for vortex-center extraction is presented in conjunction with the results from the examples that follow.


\section{Examples of vortex-center extraction}
\label{sec:examples}
Four examples are considered.
The Euler-flow and Stokes-flow examples are the same as those considered by \citet{haller2005objective}.
These two examples are used here only as test flows, and the objectivity with respect to a rotating reference frame, which was the main concern of \citet{haller2005objective}, is not considered in the present study.
The reason is that, in principle, an observer can identify a rotating reference frame by detecting the rotation of the frame itself; therefore, invariance under the rotation of the reference frame is not regarded as an essential requirement here.
The other two examples are laminar-to-turbulent transition in channel flow at \(Re_b=U_bh/\nu=5333.3\) and turbulent channel flow at \(Re_\tau=u_\tau h/\nu=1000\), where \(U_b\) is the bulk velocity, \(u_\tau\) is the friction velocity, and \(h\) is the channel half-height.
For the Euler-, Stokes-, and channel-flow examples, the domain of interest used to compute the spatially averaged acceleration in the acceleration-based method is taken to be the entire flow domain.
However, because the spatially averaged acceleration over the entire domain is zero in all these examples, no subtraction is actually required.

\subsection{Example 1: Euler flow (ABC flow)} 

\begin{figure}
\centering
\begin{subfigure}{0.32\textwidth}
  \centering
  \includegraphics[width=\linewidth,trim={0cm 0cm 0cm 0cm},clip]{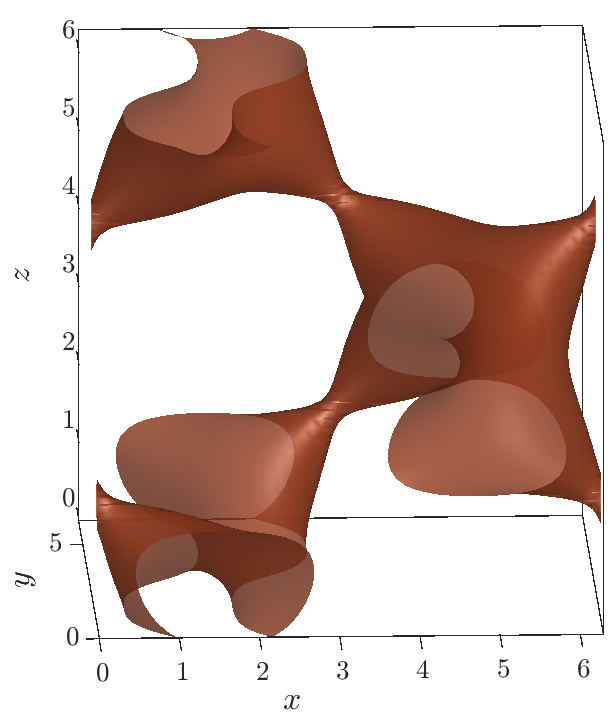}
  \caption{Isosurface of $Q_{\mathrm{sp}}=0.733$}
\end{subfigure}
\hspace{50pt}
\begin{subfigure}{0.32\textwidth}
  \centering
  \includegraphics[width=\linewidth,trim={0cm 0cm 0cm 0cm},clip]{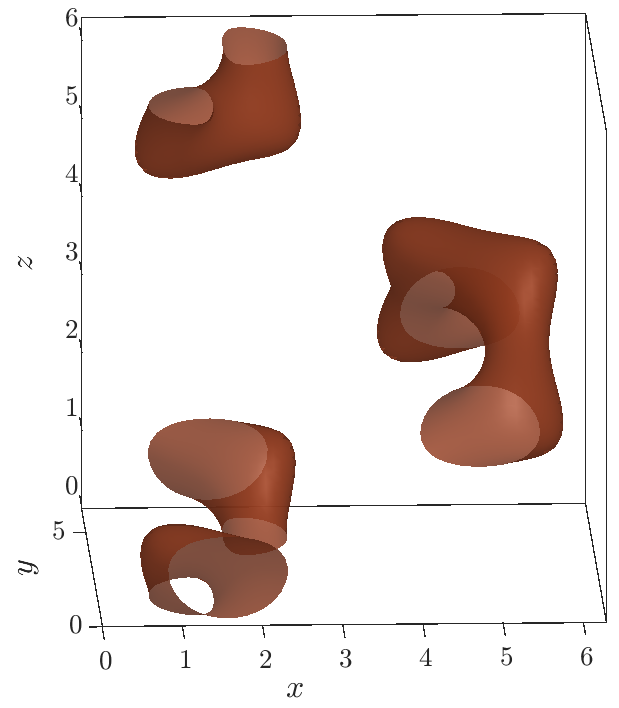}
  \caption{Isosurface of $Q_{\mathrm{sp}}=0.866$}
\end{subfigure}
\hspace{30pt}
\par\medskip

\begin{subfigure}{0.43\textwidth}
  \centering
  \includegraphics[width=\linewidth]{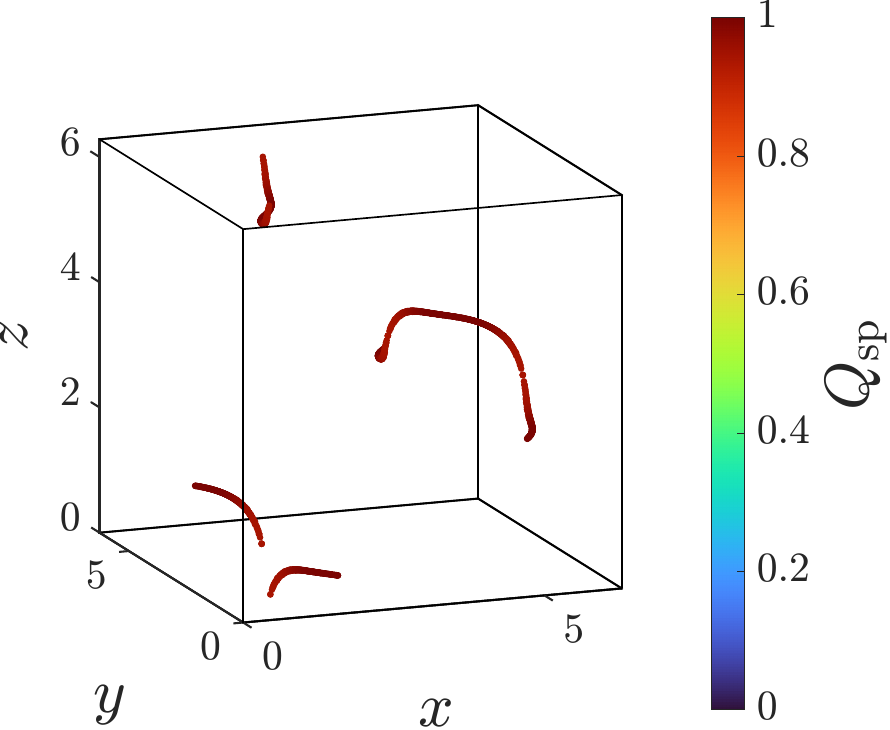}
  \caption{Vortex-strength-based vortex center \\ (local maximum of $Q_{\mathrm{sp}}$)}
\end{subfigure}
\hspace{10pt}
\begin{subfigure}{0.43\textwidth}
  \centering
  \includegraphics[width=\linewidth]{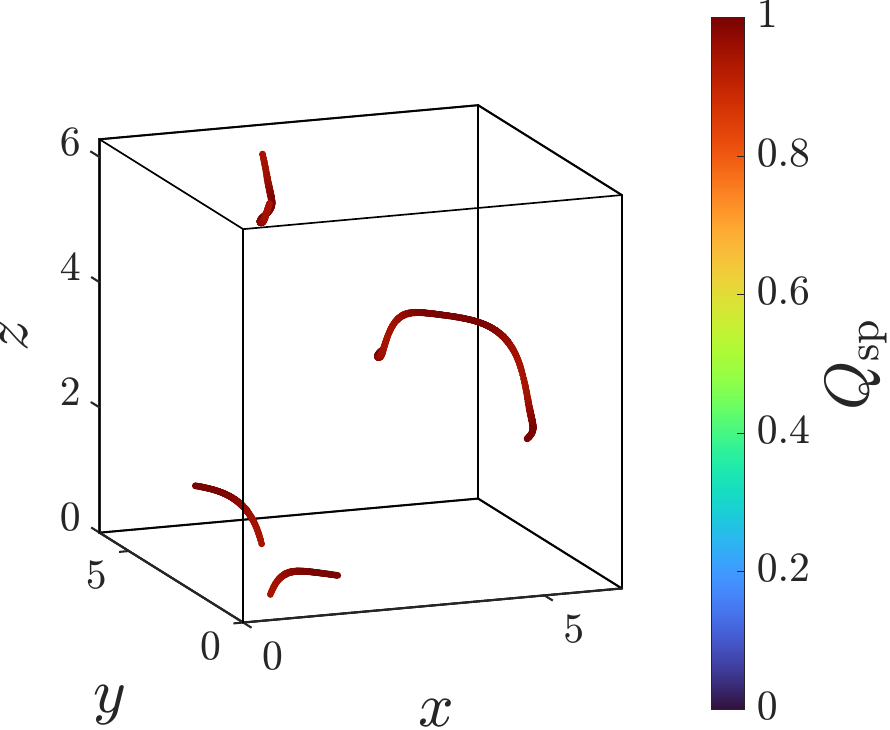}
  \caption{Acceleration-based vortex center \\ (local pressure minimum)}
\end{subfigure}
\caption{(a,b) Isosurfaces of \(Q_{\mathrm{sp}}\) at two threshold values.
(c,d) Vortex centers obtained using the vortex-strength-based and acceleration-based methods, respectively.
In (c,d), only centers with estimated radii smaller than twice the grid spacing are plotted.
The grid resolution for all four visualizations is \((n_x,n_y,n_z)=(160,160,160)\).
}
\label{fig:abc1}
\end{figure}

The ABC flow \citep{dombre1986chaotic} is considered.
The velocity field is given by
\begin{equation}
\begin{aligned}
u = A\sin z + C\cos y, \ \ 
v = B\sin x + A\cos z, \ \
w = C\sin y + B\cos x.
\end{aligned}
\label{eq:abc_flow}
\end{equation}
The parameters are set to $A=1$, $B=\sqrt{2/3}$, and $C=\sqrt{1/3}$.
The flow field is evaluated on a uniform grid of \(160^3\) points over the triply periodic domain \([0,2\pi]^3\).
Figures~\ref{fig:abc1}(a) and (b) show isosurfaces of the $Q_{\mathrm{sp}}$ criterion.
These results clearly reveal the dependence of isosurface-based vortex visualization on the selected threshold.
As the threshold increases, the connected region disappears, making it unclear whether this region represents a vortex core.
Separately, the two vortex-center extraction methods yield clear and identical vortex centerlines, as shown in figures~\ref{fig:abc1}(c) and (d).
For both methods, only centers with estimated radii smaller than twice the grid spacing, \(2\Delta x\), are plotted.
Note that, because this is an Euler flow, a zero-acceleration point satisfying the prescribed AGT condition corresponds directly to a local pressure minimum.

\begin{figure}
\captionsetup{justification=centering}
\centering
\begin{subfigure}{0.33\textwidth}
  \centering
  \includegraphics[width=\linewidth]{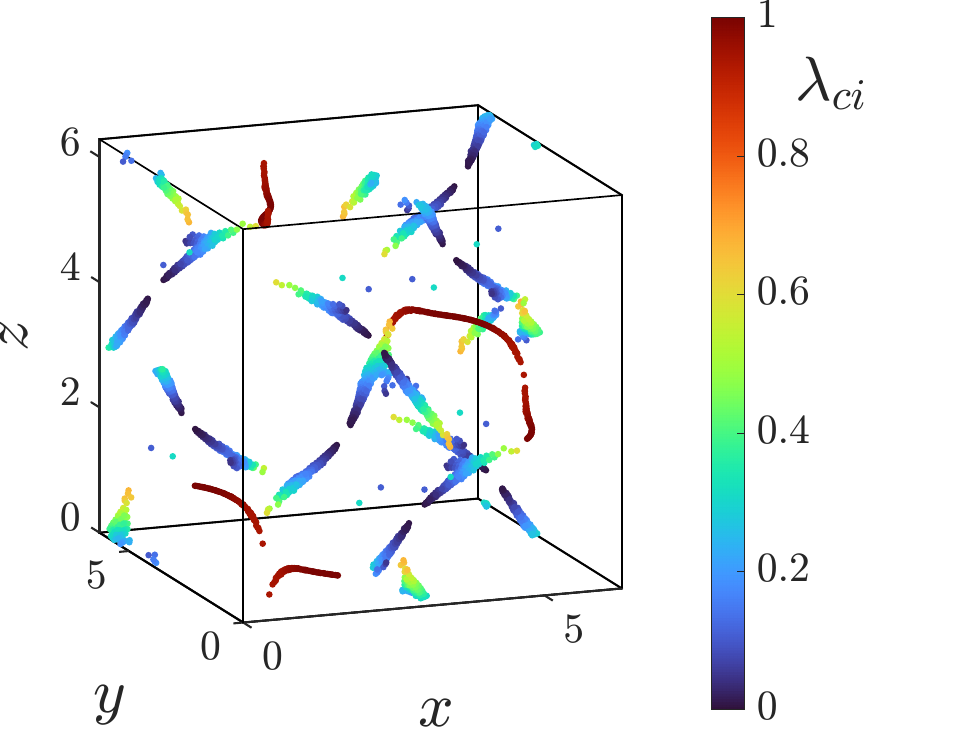}
  \caption{Local maximum of $\lambda_{ci}$}
  \vspace{\baselineskip}
\end{subfigure}%
\hfill
\begin{subfigure}{0.33\textwidth}
  \centering
  \includegraphics[width=\linewidth]{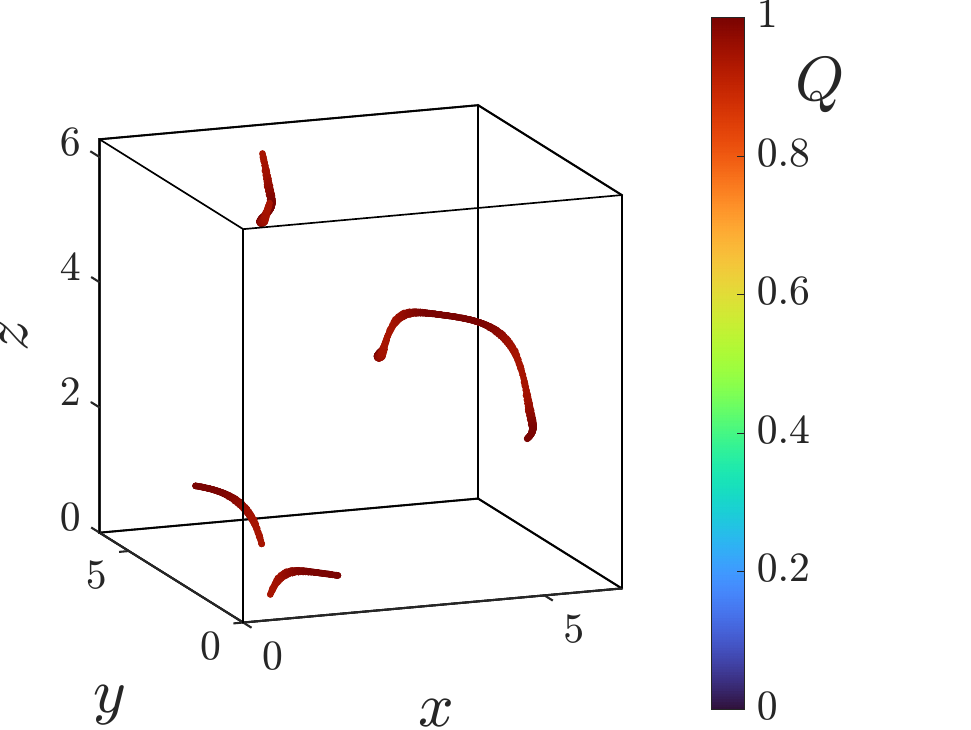}
  \caption{Local maximum of $Q$}
  \vspace{\baselineskip}
\end{subfigure}%
\hfill
\begin{subfigure}{0.33\textwidth}
  \centering
  \includegraphics[width=\linewidth]{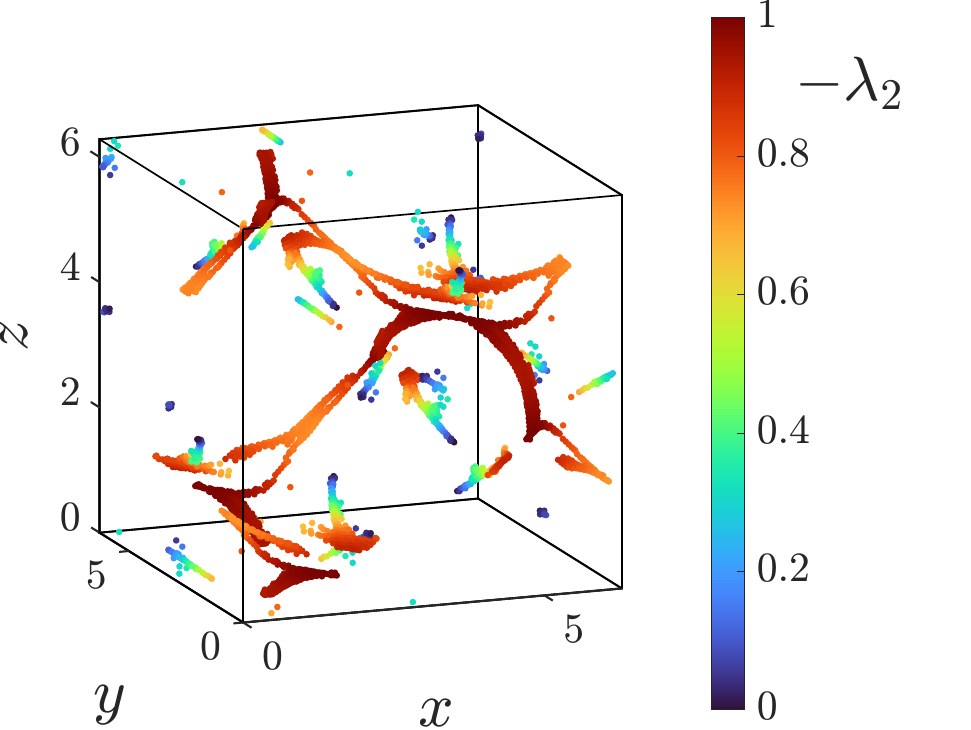}
  \caption{Local maximum of $-\lambda_2$ \\ with the associated axis}
\end{subfigure}

\par\medskip
\begin{subfigure}{0.33\textwidth}
  \centering
  \includegraphics[width=\linewidth]{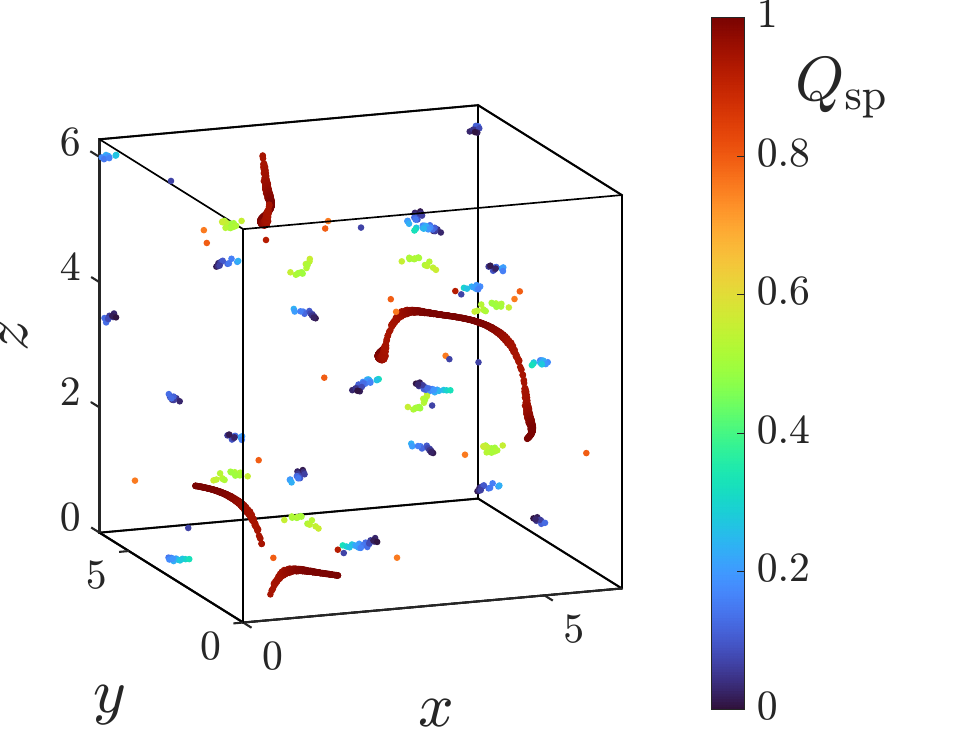}
  \caption{Local maximum of $Q_{\mathrm{sp}}$ \\ with left real eigenvector as axis}
\end{subfigure}%
\hspace{10pt}
\begin{subfigure}{0.33\textwidth}
  \centering
  \includegraphics[width=\linewidth]{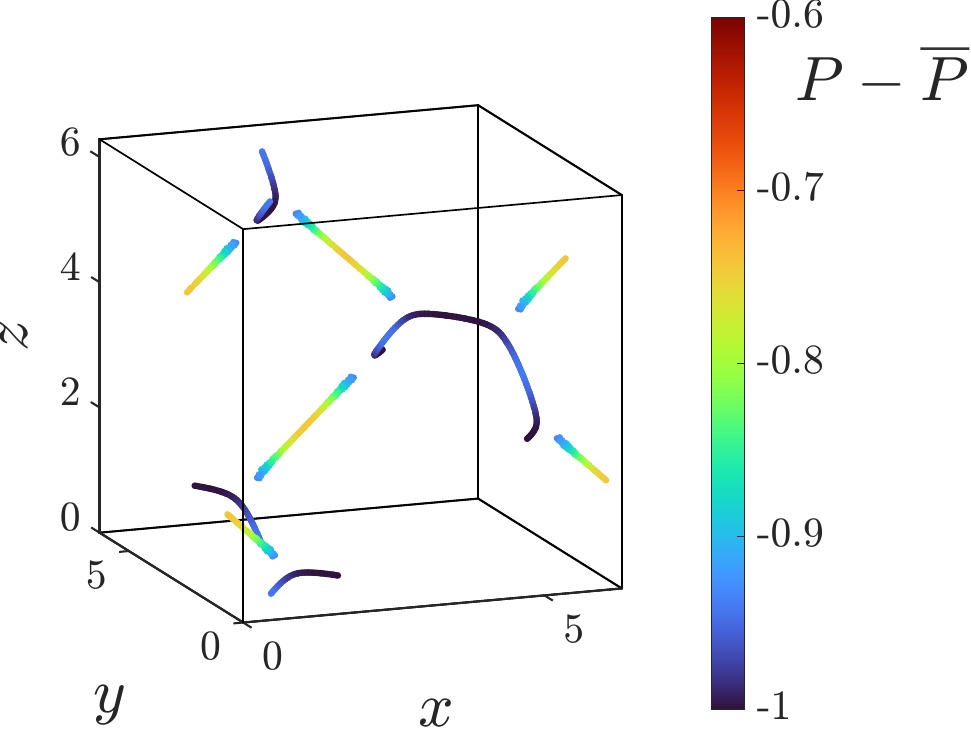}
  \caption{Local minimum of pressure \\ with the associated axis}
\end{subfigure}%

\captionsetup{justification=justified}
\caption{Vortex centers obtained using various methods, with the centers identified from the local extrema specified below.
Unless otherwise specified, the real eigenvector of the VGT is used as the axis; where an alternative axis is specified, the axis associated with the corresponding method is used.
Only centers with estimated radii smaller than twice the grid spacing are plotted.
The grid resolution is \((n_x,n_y,n_z)=(160,160,160)\).}
\label{fig:abc2}
\end{figure}

To compare the original vortex-identification criteria, \(\lambda_{ci}\), \(Q\), and \(\lambda_2\), alternative vortex-center extraction methods are constructed based on these criteria.
The alternative vortex-strength-based vortex-center methods tested here are as follows:
\begin{enumerate}
\item \textit{\(\;\lambda_{ci}\)}
\par
For the condition \textit{\(\lambda_{ci}\)}, the vortex axis is chosen in the same manner as for \(Q_{\mathrm{sp}}\), while the magnitude is defined by \(\lambda_{ci}\) (figure~\ref{fig:abc2}(a)).

\item \textit{\(\;Q\)}
\par
For the condition \textit{\(Q\)}, the vortex axis is chosen in the same manner as for \(Q_{\mathrm{sp}}\), while the magnitude is defined by \(Q\) (figure~\ref{fig:abc2}(b)).

\item \textit{\(\;\lambda_2\)}
\par
For the condition \textit{\(\lambda_2\)}, the vortex axis is chosen as the eigenvector corresponding to the largest eigenvalue, \(\lambda_1\), of \(\mathsfbi{S}^2+\mathsfbi{\Omega}^2\), while the magnitude is defined by \(-\lambda_2\) (figure~\ref{fig:abc2}(c)).

\item \textit{\(\;Q_{\mathrm{sp}}\) with the axis replaced by the real left eigenvector}
\par
In this case, the real left eigenvector of the VGT is used as the axis instead of the real right eigenvector (figure~\ref{fig:abc2}(d)).
\end{enumerate}
For the methods based on \(\lambda_{ci}\) and \(Q\), the axis is chosen in the same manner as for \(Q_{\mathrm{sp}}\) because their original formulations do not explicitly define an axis suitable for the present procedure.
For comparison with the vortex-center extraction method by \citet{miura1997identification}, their method, which belongs to the category of acceleration-based vortex-center methods, is also reproduced:
\begin{enumerate}
\setcounter{enumi}{4}
\item \textit{ Pressure-minimum} 
\par
For the pressure-minimum method, The vortex axis is chosen as the eigenvector corresponding to the largest eigenvalue of the negative pressure Hessian, while the zero-acceleration point is replaced by a pressure minimum.
This method follows \citet{miura1997identification} (figure~\ref{fig:abc2}(e)).
\end{enumerate}

The \(\lambda_{ci}\)-based method shown in figure~\ref{fig:abc2}(a) identifies additional vortex centers compared with the \(Q_{\mathrm{sp}}\)-based method shown in figure~\ref{fig:abc1}(c).
However, these centers do not cluster along a line but instead form a cone-shaped structure.
The results obtained using the \(Q\)-based method, shown in figure~\ref{fig:abc2}(b), are identical to those obtained using the \(Q_{\mathrm{sp}}\)-based method.
Therefore, no difference is observed between the local maxima identified by the \(Q\)- and \(Q_{\mathrm{sp}}\)-based methods.
In figure~\ref{fig:abc2}(d), where the real left eigenvector of the VGT is used as the axis, additional vortex centers appear either as short lines or as isolated points.
As discussed previously, this result provides an example in which using the real left eigenvector produces spurious vortex centers.

In the \(\lambda_2\)-based method shown in figure~\ref{fig:abc2}(c), additional vortex centerlines are connected to the main vortex centerlines.
The structures associated with these additional lines also appear in the low-threshold isosurface shown in figure~\ref{fig:abc1}(a).
Similar lines corresponding to these isosurface structures are more clearly observed in figure~\ref{fig:abc2}(e), where the axis is determined from the pressure Hessian following the method of \citet{miura1997identification}.
The similarity between the pressure-based and \(\lambda_2\)-based vortex centerlines can be explained by the fact that their axes are determined from the pressure Hessian and the modified pressure Hessian, \(\mathsfbi{S}^2+\mathsfbi{\Omega}^2\), respectively.
However, these centerlines are not identified using the \(Q_{\mathrm{sp}}\)-based method, for which the real eigenvector of the VGT is used as the axis, because a visualization of the two axis fields (not shown) indicates that the VGT- and pressure-Hessian-based axes are nearly perpendicular along these lines.
This result indicates that a pressure minimum does not necessarily lie within the swirling plane, providing another example of the limitation of the \(\lambda_2\) method discussed by \citet{chakraborty2005relationships}.
Furthermore, the pressure-Hessian-based axis used in figure~\ref{fig:abc2}(e) demonstrates that, in the acceleration-based method, defining the axis using an eigenvector of the AGT rather than the real eigenvector of the squared VGT can produce inappropriate vortex centerlines.
This is because, in an Euler flow, the AGT and the pressure Hessian have the same eigenvectors.

In summary, \(Q_{\mathrm{sp}}\) yields results that are clearly different from those obtained using \(\lambda_{ci}\) and \(\lambda_2\) in this ABC-flow example.
The local-maximum centerlines obtained using \(Q_{\mathrm{sp}}\) are consistent with those obtained using \(Q\) in this Euler-flow example.

\subsection{Example 2: Stokes flow}

Stokes flow provides a useful example for examining the superposition of two vortices because its linearity allows rigid-body rotation to be added while preserving a solution under compatible boundary conditions.
This example provides insight into how two vortices affect each other.
The example introduced by \citet{stone1991chaotic} is considered.
In this example, a spherical drop is immersed in an external flow combining straining motion and rigid-body rotation, as illustrated in figure~1 of \citet{stone1991chaotic}.
With the strain parameter \(a\) from \citet{stone1991chaotic} set to \(a=1\), the velocity field becomes
\begin{equation}
\begin{aligned}
u(\mathbf{x}) &= \frac{3x}{4}\left(x^2+y^2+3z^2-1\right)+\frac{1}{2}\left(\omega_y z-\omega_z y\right),\\
v(\mathbf{x}) &= \frac{3y}{4}\left(x^2+y^2+3z^2-1\right)+\frac{1}{2}\left(\omega_z x-\omega_x z\right),\\
w(\mathbf{x}) &= \frac{3z}{2}\left(1-2x^2-2y^2-z^2\right)+\frac{1}{2}\left(\omega_x y-\omega_y x\right).
\end{aligned}
\end{equation}

\begin{figure}
\centering
  \begin{subfigure}{0.30\textwidth}
  \centering
    \includegraphics[width=\linewidth,trim={0cm 0.cm -0.9cm 0.cm},clip]{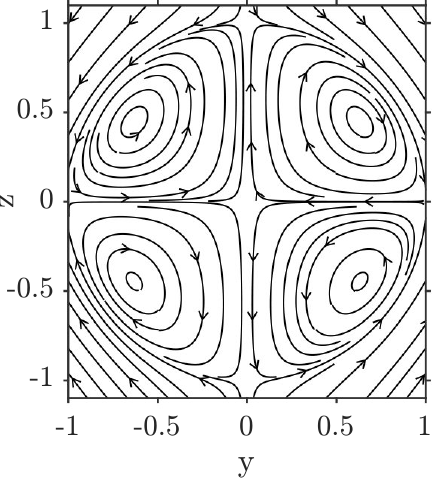}
    \caption{Streamline}
  \end{subfigure}%
  \hspace{0.08\textwidth}
  \begin{subfigure}{0.35\textwidth}
  \centering
    \includegraphics[width=\linewidth,trim={-0.6cm 0.cm 0cm 0.cm},clip]{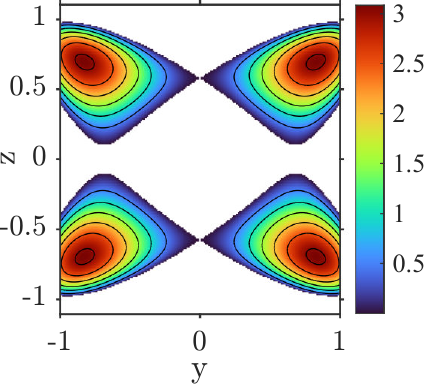}
    \caption{$Q$ contours}
  \end{subfigure}%
\caption{
Stokes-flow example of \citet{stone1991chaotic} (see figure~1 of their paper for a schematic of the flow).
(a) Streamlines in the \(y\)--\(z\) plane for \((\omega_x,\omega_y,\omega_z)=(0,0,0)\).
(b) Contours of the positive values of \(Q\) under the same conditions.
}
\label{fig:stokes1}
\end{figure}

In this incompressible flow case, the pressure field is given by $p(x,y,z)=\frac{21}{4}\mu\left(x^2+y^2-2z^2\right)$.
Consistent with the limitation of pressure-based vortex identification pointed out by \citet{jeong1995identification}, the in-plane pressure minimum does not correspond to a vortex core in this Stokes flow.
Although the pressure reaches an in-plane minimum on the \(z\)-axis when \((\omega_x,\omega_y,\omega_z)=(0,0,0)\), no vortex exists along this axis.
Moreover, the definition \(Q=\frac{1}{2}\left(\lVert\mathsfbi{\Omega}\rVert^2-\lVert\mathsfbi{S}\rVert^2\right)\) yields the distribution shown in figure~\ref{fig:stokes1}(b), which differs from the result inferred from the pressure-based relation, \(Q=\nabla^2p/(2\rho)=0\).
This apparent discrepancy arises because the convection term is neglected in the Stokes equation; consequently, the relation \(Q=\nabla^2p/(2\rho)\), derived from the full incompressible momentum equation, cannot be used to evaluate \(Q\) in this Stokes-flow example.

\begin{figure}
\captionsetup{justification=centering}
\centering
  \begin{subfigure}[t]{0.32\textwidth}
  \centering
    \includegraphics[width=\linewidth,trim={0cm 0.cm 0cm 0.cm},clip]{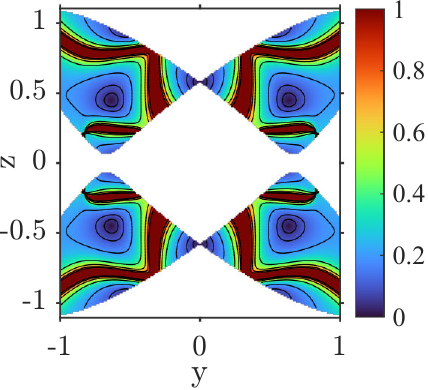}
    \caption{Acceleration based $|\boldsymbol{r}|$}
  \end{subfigure}%
  \hspace*{\fill}   
  \begin{subfigure}[t]{0.32\textwidth}
  \centering
    \includegraphics[width=\linewidth,trim={0cm 0.cm 0cm 0.cm},clip]{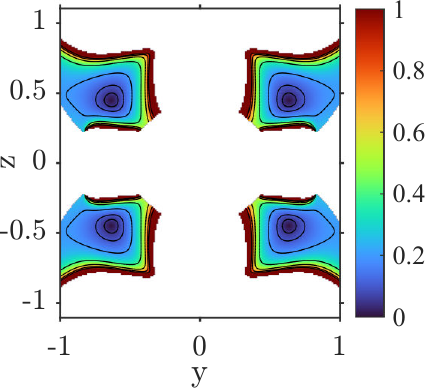}
    \caption{Acceleration-based $|\boldsymbol{r}|$ satisfying the prescribed condition}
  \end{subfigure}%
  \hspace*{\fill}   
  \begin{subfigure}[t]{0.32\textwidth}
  \centering
    \includegraphics[width=\linewidth,trim={0cm 0.cm 0cm 0.cm},clip]{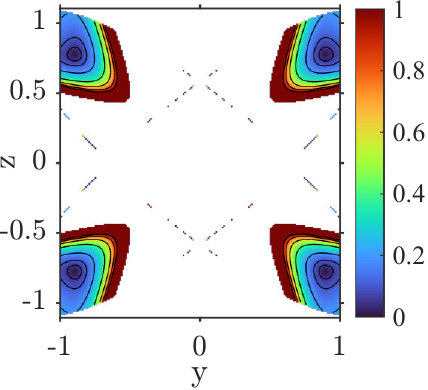}
    \caption{Vortex strength based $|\boldsymbol{r}|$ satisfying the prescribed condition}
  \end{subfigure}%
  \captionsetup{justification=justified}
\caption{Results for \((\omega_x,\omega_y,\omega_z)=(0,0,0)\).
(a) Estimated distances \(|\boldsymbol{r}|\) to the zero-acceleration point for data satisfying \(Q_{\mathrm{sp}}>0\).
(b) Estimated distances to the zero-acceleration point with the additional prescribed AGT condition imposed.
(c) Estimated distances to the local maximum of \(Q_{\mathrm{sp}}>0\) for data satisfying the prescribed condition on the Hessian of \(Q_{\mathrm{sp}}\).
In (a) and (b), the deep-blue region, where \(|\boldsymbol{r}|\to0\), represents the vortex center identified by the acceleration-based method.}
\label{fig:stokes2}
\end{figure}

The estimated radius obtained using the acceleration-based method in the \(y\)--\(z\) plane is shown in figure~\ref{fig:stokes2}(a).
Because the flow pattern in this Stokes-flow example is relatively simple and symmetric, the vortex centers are identified directly from contours of the estimated radius rather than by plotting the estimated center positions.
Only data satisfying \(Q_{\mathrm{sp}}>0\) are plotted in (a), with the remaining regions shown in white.
Among these data, only those that also satisfy the previously prescribed AGT condition, thereby fully satisfying the conditions of the acceleration-based method, are shown in figure~\ref{fig:stokes2}(b).
Note that multiple candidate vortex centers appear in each quadrant in (a), whereas only one vortex center is identified in each quadrant in (b), consistent with the intuitive centers indicated by the streamline pattern.
However, the vortex centers identified using the vortex-strength-based method based on \(Q_{\mathrm{sp}}\), shown in figure~\ref{fig:stokes2}(c), are located farther from the origin of the \(y\)--\(z\) plane than those identified using the acceleration-based method.

\begin{figure}
\centering
  \begin{subfigure}{0.32\textwidth}
    \centering
    \includegraphics[width=\linewidth,trim={0cm 0.cm 0cm 0.cm},clip]{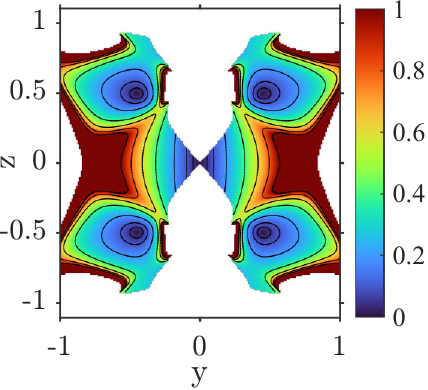}
    \caption{$\omega_z = 1.5$} \label{fig:top_a}
  \end{subfigure}%
  \hfill
  \begin{subfigure}{0.32\textwidth}
    \centering
    \includegraphics[width=\linewidth,trim={0cm 0.cm 0cm 0.cm},clip]{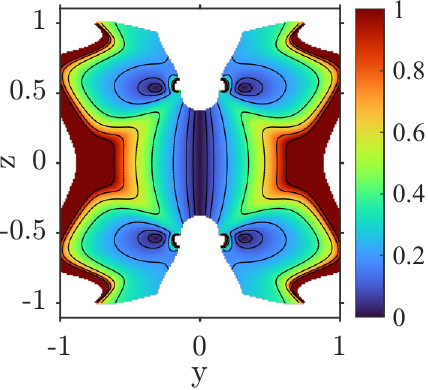}
    \caption{$\omega_z = 2$} \label{fig:top_b}
  \end{subfigure}%
  \hfill
  \begin{subfigure}{0.32\textwidth}
    \centering
    \includegraphics[width=\linewidth,trim={0cm 0.cm 0cm 0.cm},clip]{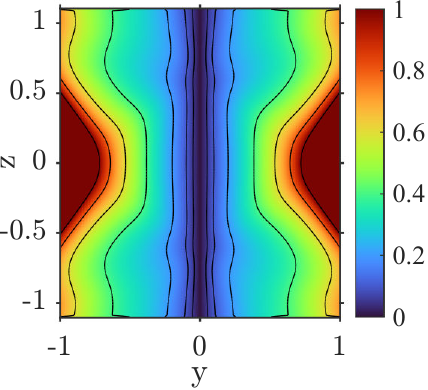}
    \caption{$\omega_z = 4$} \label{fig:top_c}
  \end{subfigure}
  \caption{Estimated distances to the zero-acceleration points obtained using the acceleration-based method for the $\omega_z=1.5$, $2$, and $4$ cases.}
 \label{fig:stokes_acc}

\begin{subfigure}[t]{0.32\textwidth}
  \vspace{0pt}
  \centering
  \includegraphics[width=\linewidth]{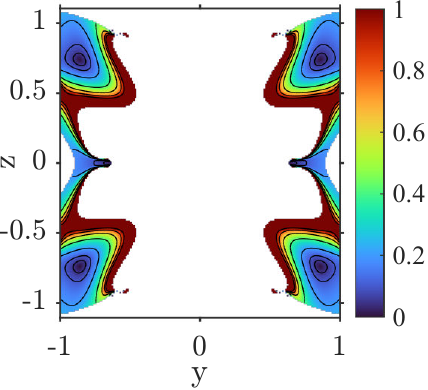}
  \caption{$\omega_z = 1.5$}\label{fig:bot_a}
\end{subfigure}\hfill%
\begin{subfigure}[t]{0.32\textwidth}
  \vspace{0pt}
  \centering
  \includegraphics[width=\linewidth]{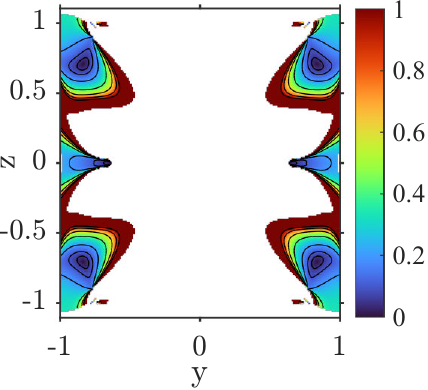}
  \caption{$\omega_z = 2$}\label{fig:bot_b}
\end{subfigure}\hfill%
\begin{subfigure}[t]{0.32\textwidth}
  \vspace{0pt}
  \centering
  \includegraphics[width=\linewidth]{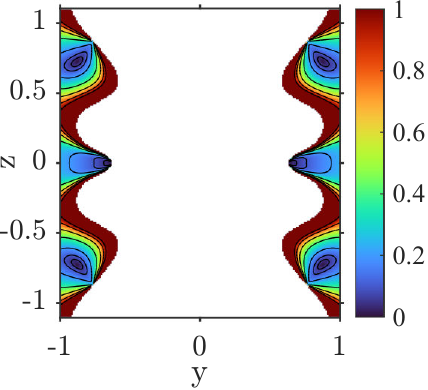}
  \caption{$\omega_z = 4$}\label{fig:bot_c}
\end{subfigure}
\captionsetup{justification=justified}
 \caption{Estimated distances to the local maxima of \(Q_{\mathrm{sp}}\)  for the $\omega_z=1.5$, $2$, and $4$ cases.}
  \label{fig:stokes_qsp}
\end{figure}

Thus far, the global vorticity has been set to \((\omega_x,\omega_y,\omega_z)=(0,0,0)\).
To examine the effect of superposing two vortical motions, a nonzero \(\omega_z\) is added.
Without Stokes-flow approximation, such a linear superposition is generally not valid.
Even in a Stokes flow, the superposition considered here is restricted to rigid-body rotation, so its effect may differ from the interaction between two ordinary vortices whose centers have the highest vortex strength. 
Figures~\ref{fig:stokes_acc} and \ref{fig:stokes_qsp} present the results obtained using the acceleration-based and vortex-strength-based methods for \(\omega_z=1.5\), \(2\), and \(4\).
The acceleration-based results in figures~\ref{fig:stokes_acc}(a)--(c) clearly exhibit a centerline along the \(z\)-axis.
In addition, as \(\omega_z\) increases, the off-axis vortex centers move toward the centerline on the \(z\)-axis and disappear in the \(\omega_z=4\) case.

In figures~\ref{fig:stokes_acc}(a)--(c), a centerline appears along the \(z\)-axis, and its axial extent increases with \(\omega_z\).
As \(\omega_z\) increases from \(1.5\) to \(2\), the centerline along the \(z\)-axis extends, while the off-axis vortex centers move toward the axis, as shown in figures~\ref{fig:stokes_acc}(a) and (b).
This behavior can be interpreted as the original off-axis vortices and the vortical motion along the \(z\)-axis jointly producing vortex centers between them, with the vortex-center positions determined by the strength of each vortical motion.
Finally, the off-axis vortex centers disappear at \(\omega_z=4\), as shown in figure~\ref{fig:stokes_acc}(c).
In this steady flow, the original and added vortical motions generate the acceleration required to maintain the fluid-particle trajectories; as the contribution of the added rigid-body rotation increases, the zero-acceleration points move toward the \(z\)-axis.
Thus, in this example, the positions of the acceleration-based vortex centers depend on the relative contributions of the interacting vortical motions to the acceleration field.

In contrast, in the vortex-strength-based results shown in figures~\ref{fig:stokes_qsp}(a)--(c), the locations of the local maxima of \(Q_{\mathrm{sp}}\) remain nearly unchanged.
This indicates that the original vortex centers persist even in the presence of the added vortical motion.
An additional vortex center is also detected in the \(z=0\) plane.
The original vortex centers are retained because the added rigid-body rotation has spatially uniform vorticity and, in this example, does not substantially shift the local maxima of \(Q_{\mathrm{sp}}\).

Because a simple linear superposition of two non-rigid vortices is not generally valid, their combined effect cannot be examined directly in the same manner.
Nevertheless, these results suggest that, when vortices interact, their relative contributions to the acceleration field determine the positions of the combined vortex centers in the acceleration-based results.
If all the other interacting vortices are much weaker, a combined vortex center would remain near the center of the strongest vortex and would therefore be associated with relatively high vortex strength.
This interpretation would be supported if, in the following transitional- and turbulent-flow examples, acceleration-based vortex centers were found to be less abundant than vortex-strength-based vortex centers and generally associated with relatively high vortex strength.

In summary, for \((\omega_x,\omega_y,\omega_z)=(0,0,0)\), the acceleration-based vortex centers coincide with the intuitive centers of rotation, whereas the vortex-strength-based centers differ substantially from them.
When a global rigid-body rotation characterized by a nonzero \(\omega_z\) is added, the acceleration-based vortex centers emerge as the centers of combined vortices, and their locations depend on the relative vortex strengths of the original and added vortical motions.
The following examples examine whether this behavior persists in more complex flows.

\subsection{Example 3: natural transition in channel flow}

\begin{figure}
\captionsetup{justification=centering}
\centering
\mulimg[width=0.512\textwidth, trim=3 0 172 11, clip]{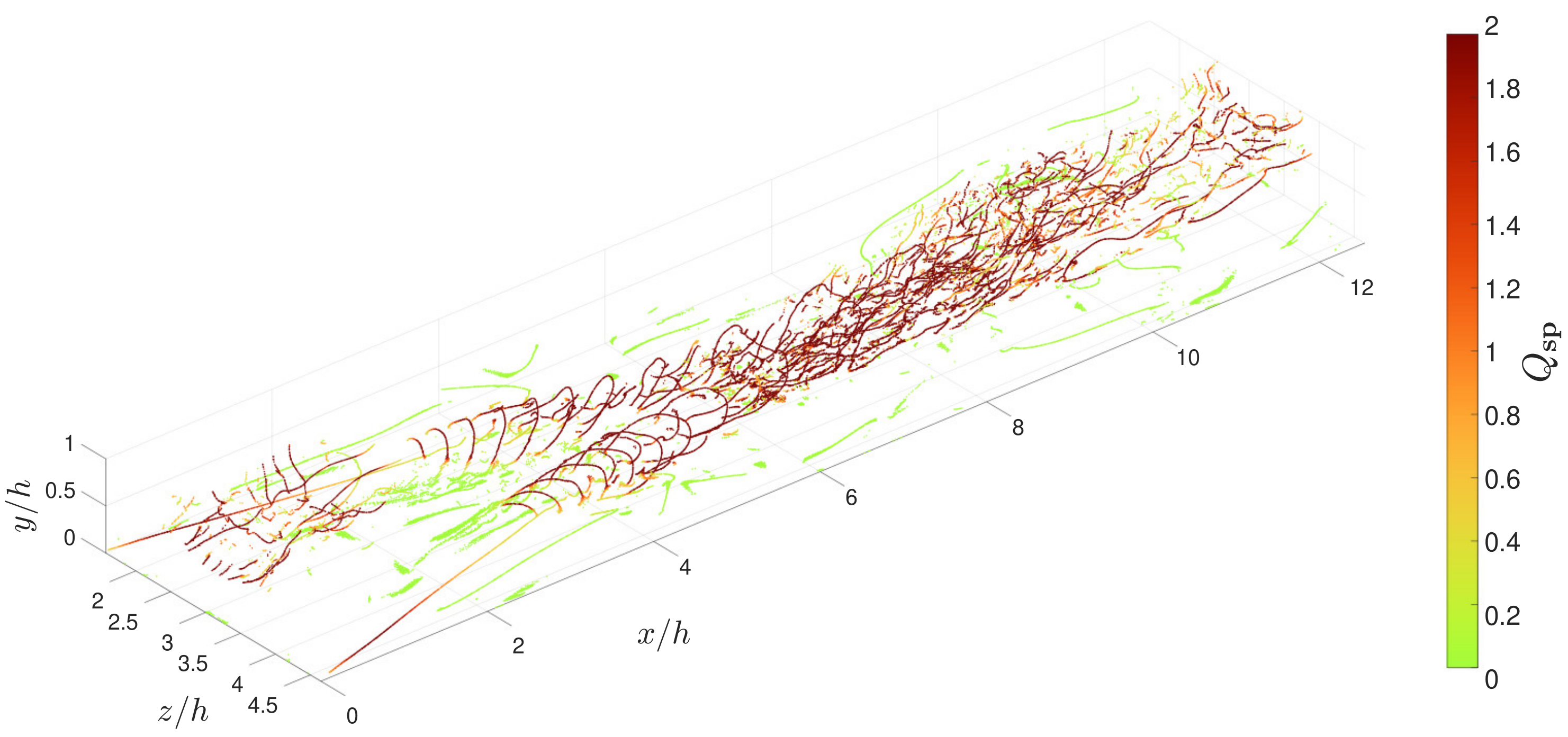}%
\hspace{-0.104\textwidth}%
\mulimg[width=0.581\textwidth, trim=3 0 5 11, clip]{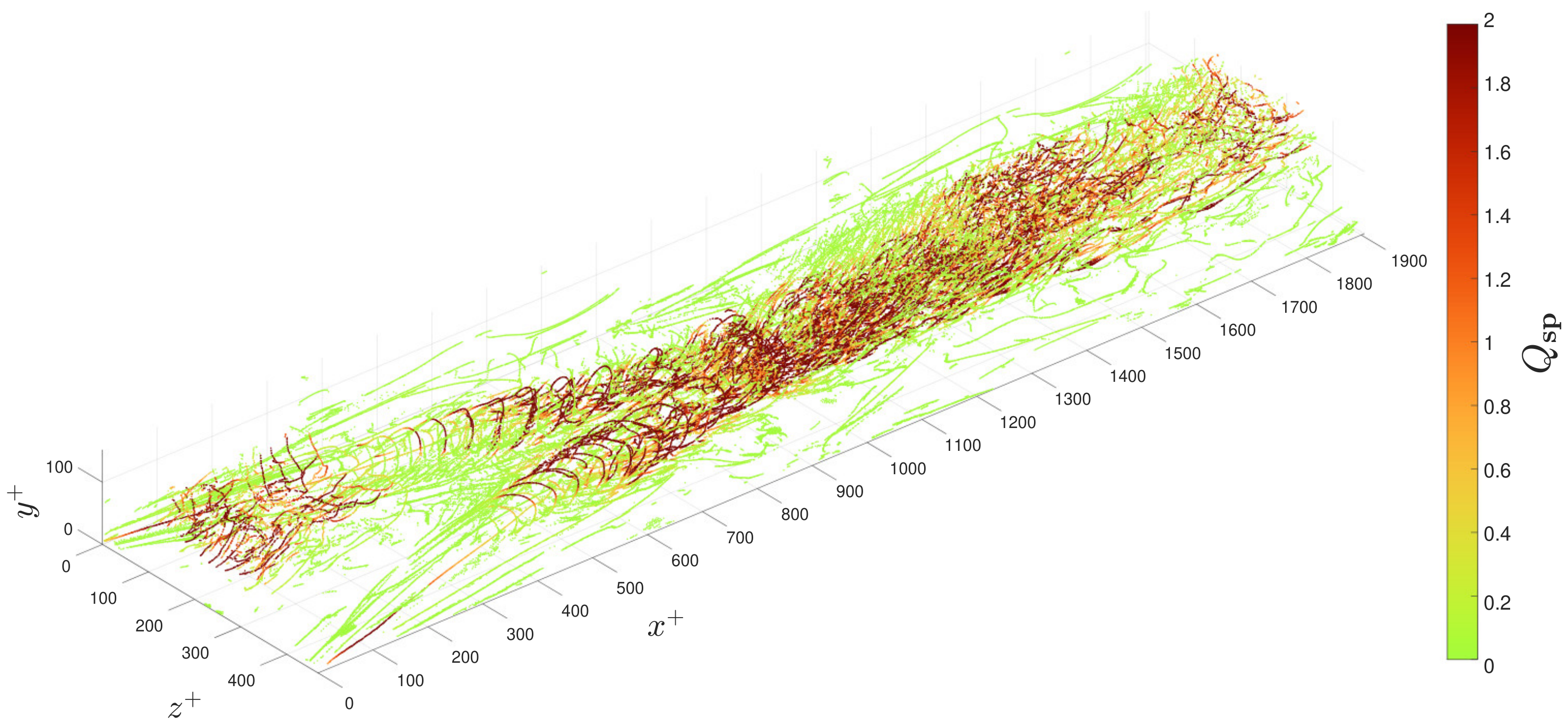}%
\par\vspace{2pt}
\begin{subfigure}[t]{0.49\textwidth}
  \caption{Acceleration-based vortex centers\protect\newline (early stage)}
\end{subfigure}\hfill
\begin{subfigure}[t]{0.49\textwidth}
  \caption{Vortex-strength-based vortex centers\protect\newline (early stage)}
\end{subfigure}

\par\medskip

\includegraphics[width=0.4465\textwidth, trim=3 0 191 0, clip]{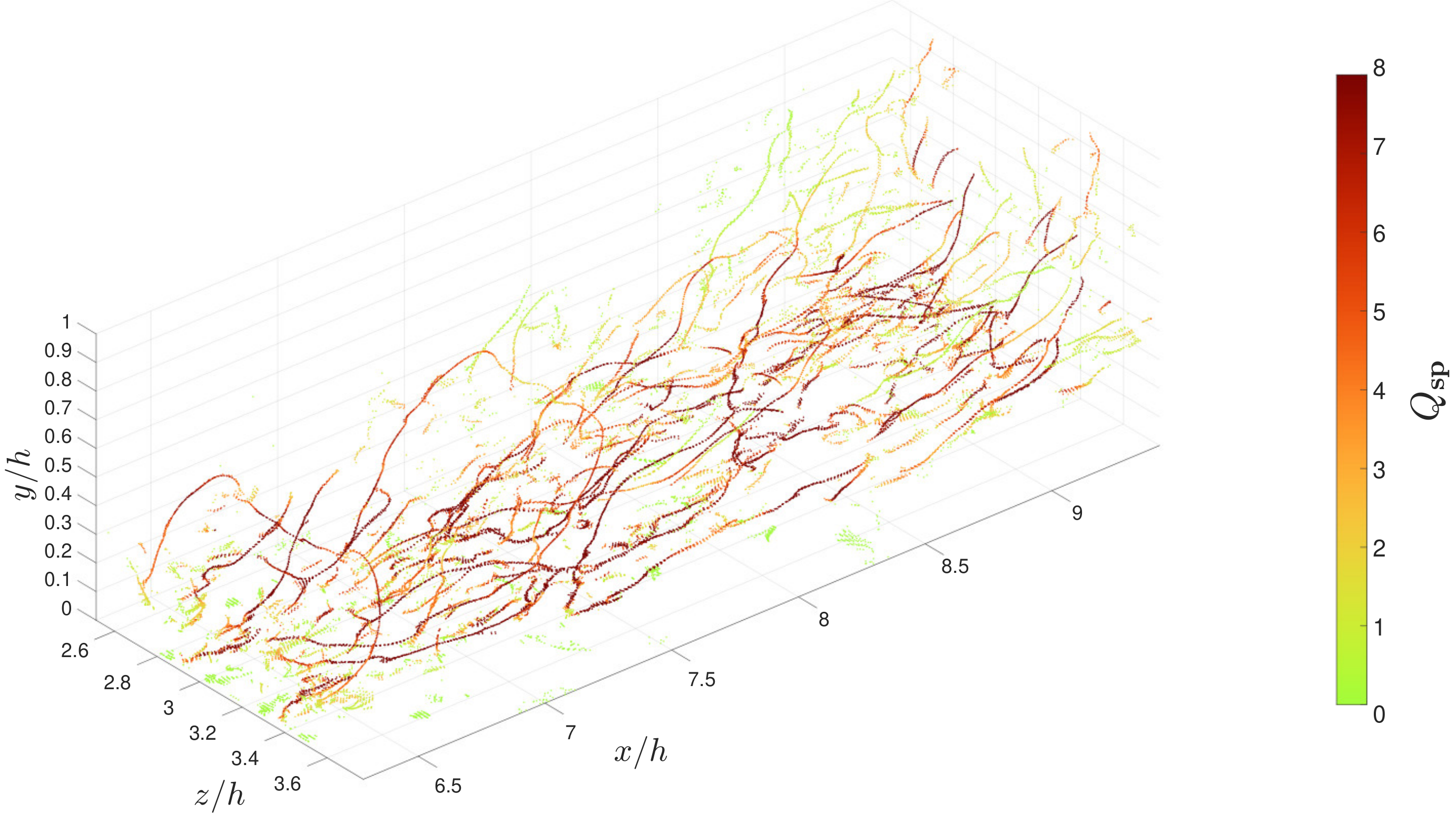}%
\hspace{0.019\textwidth}%
\includegraphics[width=0.5255\textwidth, trim=3 0 4 0, clip]{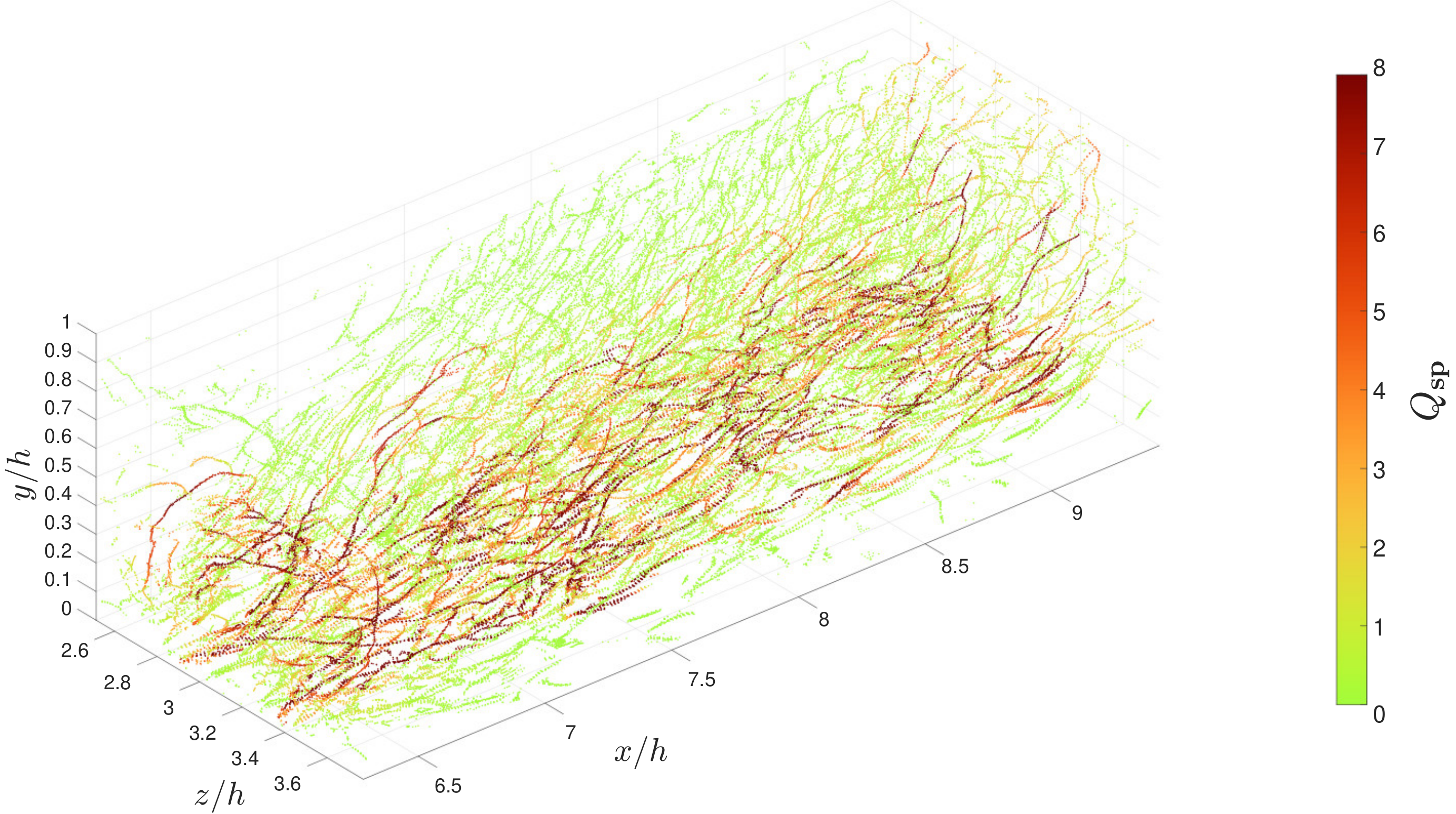}%
\par\vspace{2pt}
\begin{subfigure}[t]{0.49\textwidth}
  \caption{Acceleration-based vortex centers, \protect\newline zoomed-in view}
\end{subfigure}\hfill
\begin{subfigure}[t]{0.49\textwidth}
  \caption{Vortex-strength-based vortex centers, \protect\newline zoomed-in view}
\end{subfigure}
\captionsetup{justification=justified}
\caption{Vortex centers during the early stage of laminar-to-turbulent transition (a,b), with zoomed-in views of (a) and (b) shown in (c) and (d), respectively.
The acceleration-based results are shown in the left column (a,c) and the vortex-strength-based results in the right column (b,d).
The vortex centers are colored by the vortex strength $Q_{\mathrm{sp}}$.}
\label{fig:tran2}
\end{figure}

\subsubsection{Computational setup}
For this transitional-flow example and the turbulent-flow example that follows, the GPU code of \citet{kim2025highly} was used.
The code solves incompressible flow on a Cartesian grid using a semi-implicit fractional-step method.
The viscous term is treated using the Crank--Nicolson method, whereas the convective term is treated using a third-order low-storage Runge--Kutta method.
Spatial discretization is performed using a second-order central finite-difference scheme on a staggered grid.
The code has been validated by \citet{ha_multi-gpu_2021} and \citet{kim2025highly}.

DNS of H-type transition in channel flow was performed by introducing an initial disturbance field obtained from the Orr--Sommerfeld and Squire equations, as in \citet{kim2025highly}.
The fourth-order Orr--Sommerfeld equation and second-order Squire equation was solved using the Chebyshev collocation method \citep{biringen1988solution,canuto2006spectral} to obtain an accurate solution.
The initial disturbance follows the configuration of \citet{lee2017simulations}, with the Weissenberg number set to zero to exclude viscoelastic effects.
It is formed by superposing three modes obtained from the Orr--Sommerfeld and Squire equations, with wavenumber pairs \((k_x,k_z)=(1,0)\), \((0.5,-1)\), and \((0.5,1)\), where \(k_x\) and \(k_z\) denote the streamwise and spanwise wavenumbers, respectively.

The initial disturbance amplitudes are set to \(0.01\), \(0.0002\), and \(0.0002\) for the respective modes.
To validate the code, the growth rate of a two-dimensional Orr--Sommerfeld mode was evaluated using the setup of \citet{lee2017simulations} in a domain of size \((L_x,L_y,L_z)=(2\pi h,2h,h)\), discretized using \((N_x,N_y,N_z)=(160,2048,16)\).
The computed growth rate differed from the theoretical value by less than \(1\,\%\).
The transition simulation was conducted at a bulk Reynolds number of \(Re_b=5333.3\) in a computational domain of \((L_x,L_y,L_z)=(4\pi h,2h,2\pi h)\), discretized using \((N_x,N_y,N_z)=(256,256,256)\).

\subsubsection{Results}
To visualize the vortex centerlines more clearly, the data are interpolated by factors of four and two in the \(x\) and \(z\) directions, respectively.
The quantities required for the calculation---namely, the pressure, velocity, and velocity derivatives \(p\), \(u_i\), \(u_{i,j}\), and \(u_{i,jj}\)---are computed on the original grid and subsequently interpolated onto a finer rectilinear grid using cubic spline interpolation.
This interpolation is necessary because the original grid spacings, \(\Delta x\) and \(\Delta z\), are relatively large.
Without interpolation, centerlines nearly aligned with either the \(x\)- or \(z\)-direction appear discontinuous and are therefore difficult to recognize because only one estimated vortex center is plotted for each grid point.
For the zoomed-in views only, the data are additionally interpolated by a factor of two in the \(y\) direction.

Figure~\ref{fig:tran2} shows the region extending from the wall to the channel center over the full streamwise extent and half the spanwise extent.
The acceleration-based and vortex-strength-based results are shown in (a) and (b), respectively.
In Figure~\ref{fig:tran2}(a), data for \(y\leq 0.0021\), corresponding to the first seven grid cells, are excluded because they form a somewhat noisy, clustered pattern far from the primary turbulent region and have very low intensity (\(Q_{\mathrm{sp}}<3\times10^{-5}\)).
Zoomed-in views of the results in (a) and (b), over \(x\in[2\pi,3\pi]\) and \(z\in[0.8\pi,1.2\pi]\), are shown in (c) and (d), respectively.
An estimated center is plotted only when its estimated radius $|\boldsymbol{r}|$ is smaller than \(0.015h\).

As shown in figures~\ref{fig:tran2}(a)--(d), most of the identified centers align to form clear centerlines.
Because of the finite grid spacing, these centerlines are actually composed of discrete center points, as can be seen upon close inspection of figures~\ref{fig:tran2}(c) and (d).
No iterative procedure or explicit line-connection algorithm was used to visualize the centers as lines; only the original data were interpolated before the centers were calculated.
Even without interpolation, the same centerline structures would remain identifiable, although their constituent centers would be more widely spaced and exhibit greater positional noise.
These observations suggest that most of the identified vortex centers exhibit filamentary structures in transitional channel flow.

Clear differences are observed between the acceleration-based and vortex-strength-based vortex centerlines.
The acceleration-based vortex centerlines are less abundant.
In particular, most vortex centerlines identified by the acceleration-based method correspond to vortex centerline identified by the vortex-strength-based method, and the vortices common to both methods exhibit relatively high vortex strength within their respective neighborhoods.
Some vortices with relatively low vortex strength are also identified by the acceleration-based method outside the primary turbulent region, apparently in locations where no strong vortices are present nearby.
Before comparing the acceleration-based and vortex-strength-based methods further, another turbulent-flow example is presented in the next section.

\subsection{Example 4: turbulent flow in $Re_\tau=1000$}

\subsubsection{Computational setup}
Turbulent channel flow at \(Re_\tau=1000\) was examined.
The simulation was conducted in a domain of size \((L_x,L_y,L_z)=(6\pi h,2h,3\pi h)\), discretized using \((N_x,N_y,N_z)=(1536,512,1536)\) grid points, where \(h\) is the channel half-height.
A hyperbolic-tangent grid distribution was used in the wall-normal direction to refine the grid near the walls.
The streamwise and spanwise grid spacings were \(\Delta x^+=12.3\) and \(\Delta z^+=6.14\), respectively, and the first wall-normal grid spacing was \(\Delta y^+=0.138\).
As in the transitional-flow case, the data were interpolated by factors of four and two in the \(x\) and \(z\) directions, respectively, to improve the visual continuity of the vortex centerlines.

\subsubsection{Results}

\begin{figure}
\captionsetup{justification=centering}
\centering
\raisebox{-0.5\height}{\includegraphics[width=0.604\textwidth]{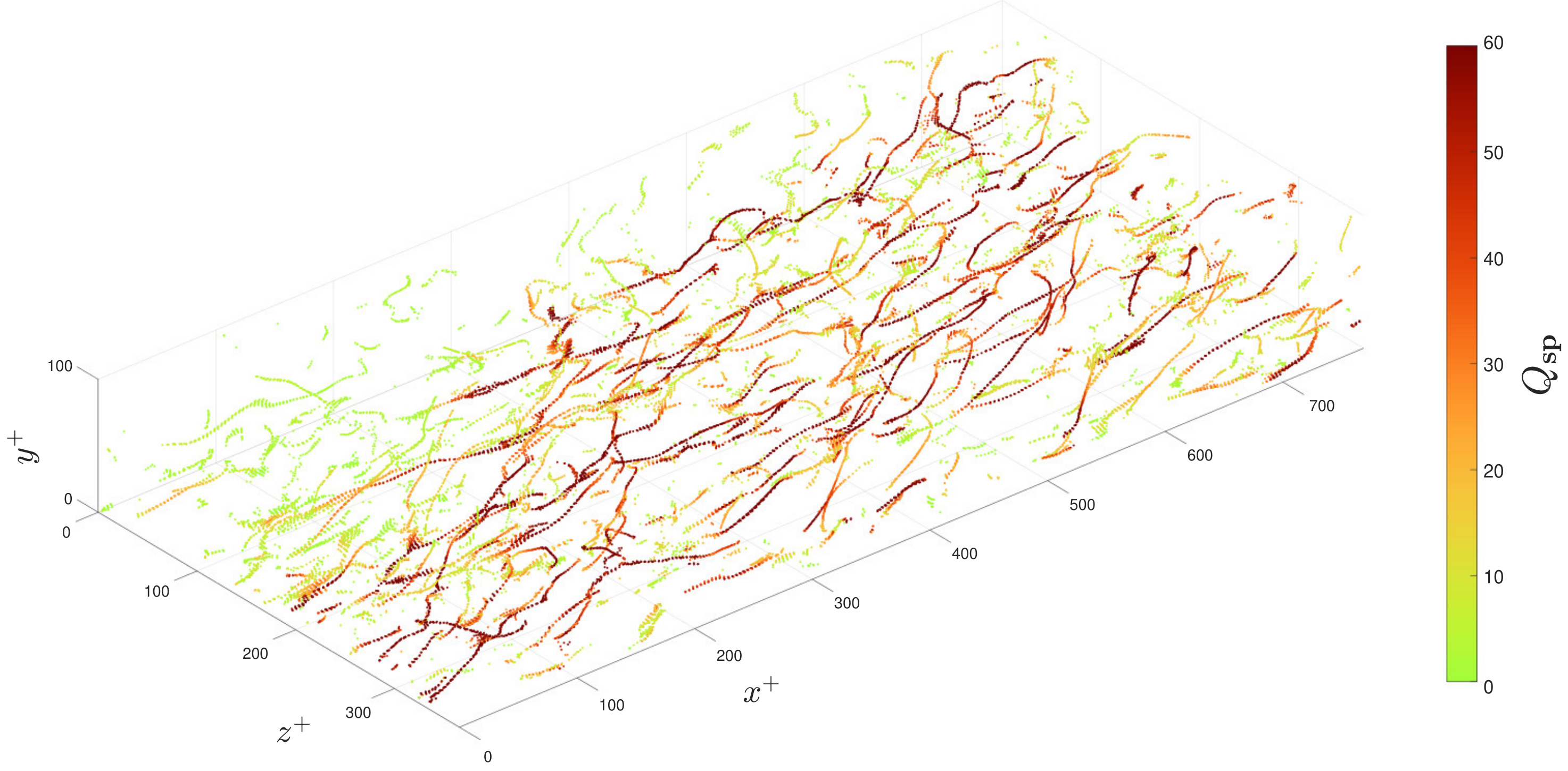}}%
\hfill
\raisebox{-0.5\height}{\includegraphics[width=0.376\textwidth]{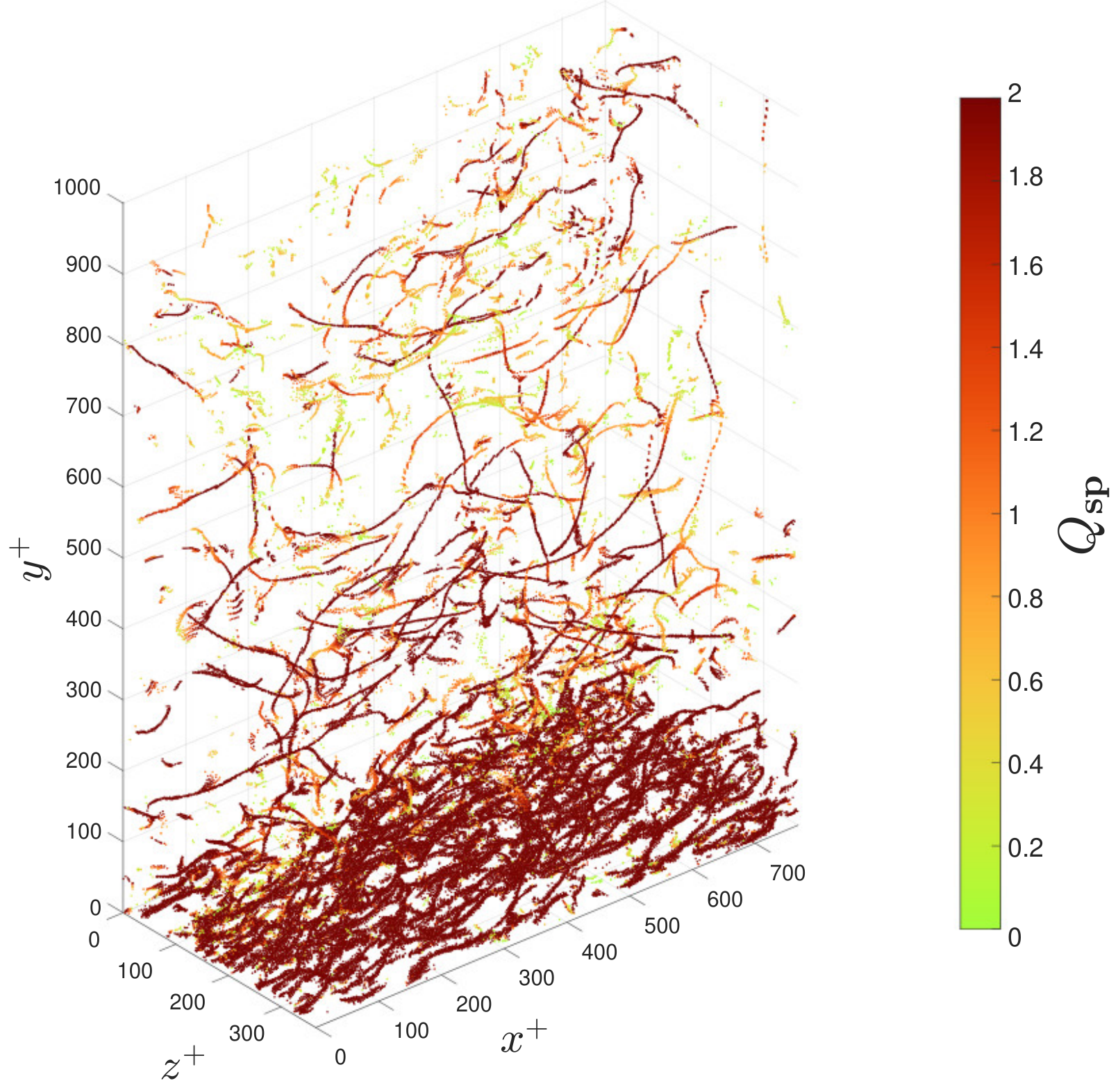}}%
\par\vspace{2pt}
\begin{subfigure}[t]{0.604\textwidth}
  \captionsetup{width=0.623\linewidth}
  \caption{Acceleration-based centers \\ ($0 \leq y^+ \leq 100$)}
\end{subfigure}
\hfill
\begin{subfigure}[t]{0.376\textwidth}
  \caption{Acceleration-based centers \\ ($0 \leq y^+ \leq 1000$)}
\end{subfigure}

\par\medskip

\raisebox{-0.5\height}{\includegraphics[width=0.604\textwidth]{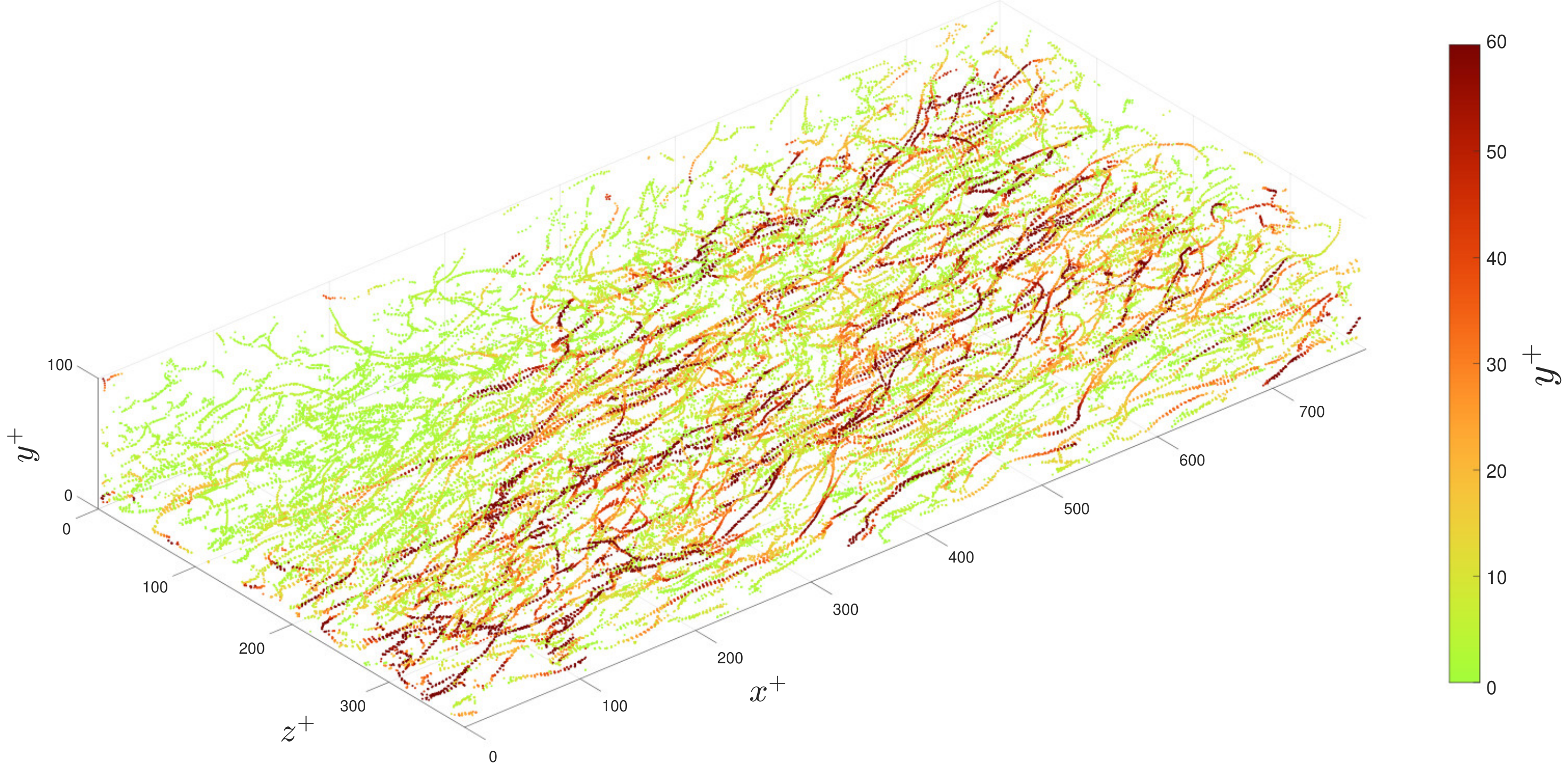}}%
\hfill
\raisebox{-0.5\height}{\includegraphics[width=0.376\textwidth]{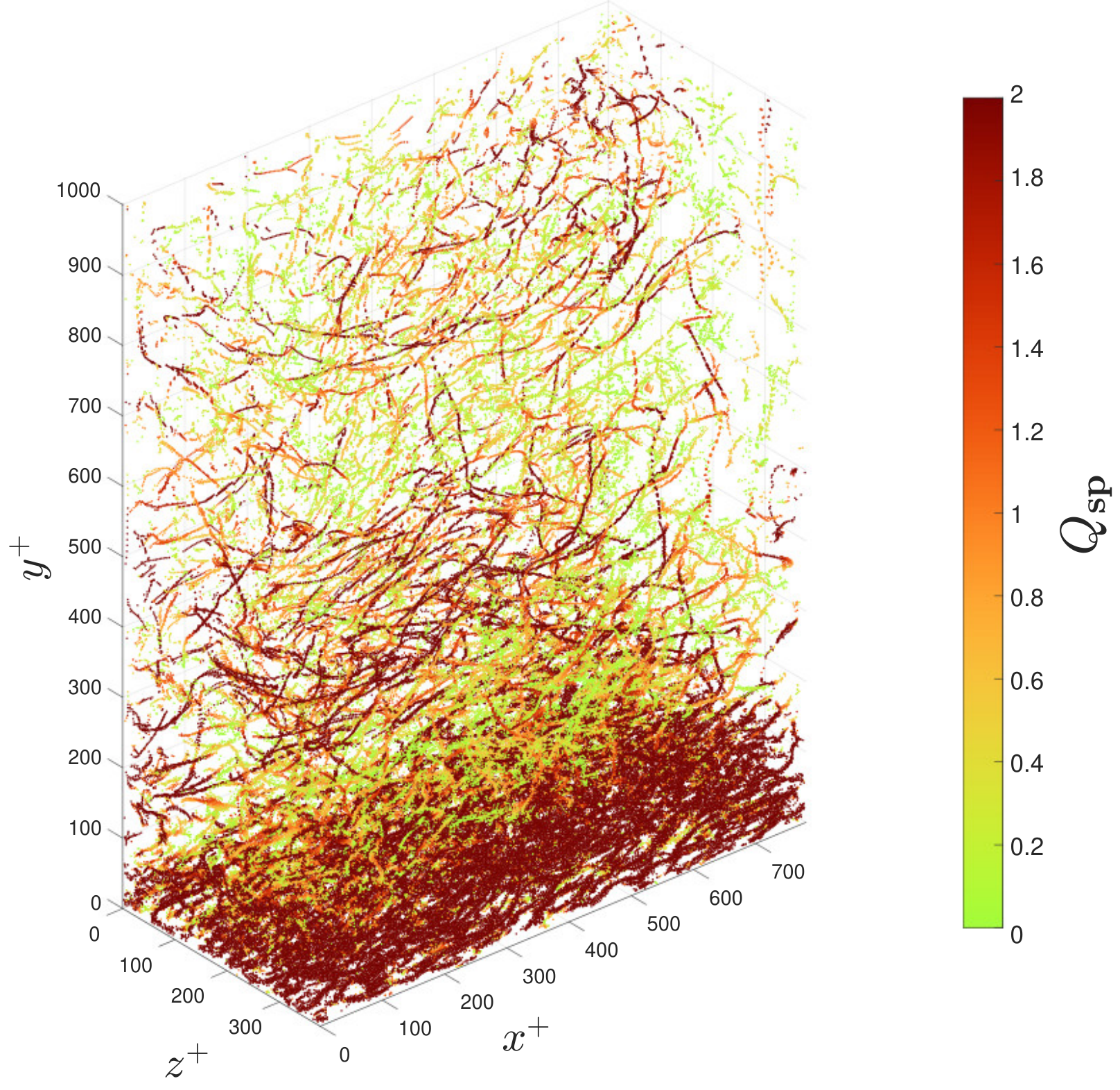}}%
\par\vspace{2pt}
\begin{subfigure}[t]{0.604\textwidth}
  \captionsetup{width=0.623\linewidth}
  \caption{Vortex-strength-based centers \\ ($0 \leq y^+ \leq 100$)}
\end{subfigure}
\hfill
\begin{subfigure}[t]{0.376\textwidth}
  \caption{Vortex-strength-based centers \\ ($0 \leq y^+ \leq 1000$)}
  \label{fig:turb1}
\end{subfigure}

\par\medskip

\raisebox{-0.5\height}{\includegraphics[width=0.606\textwidth]{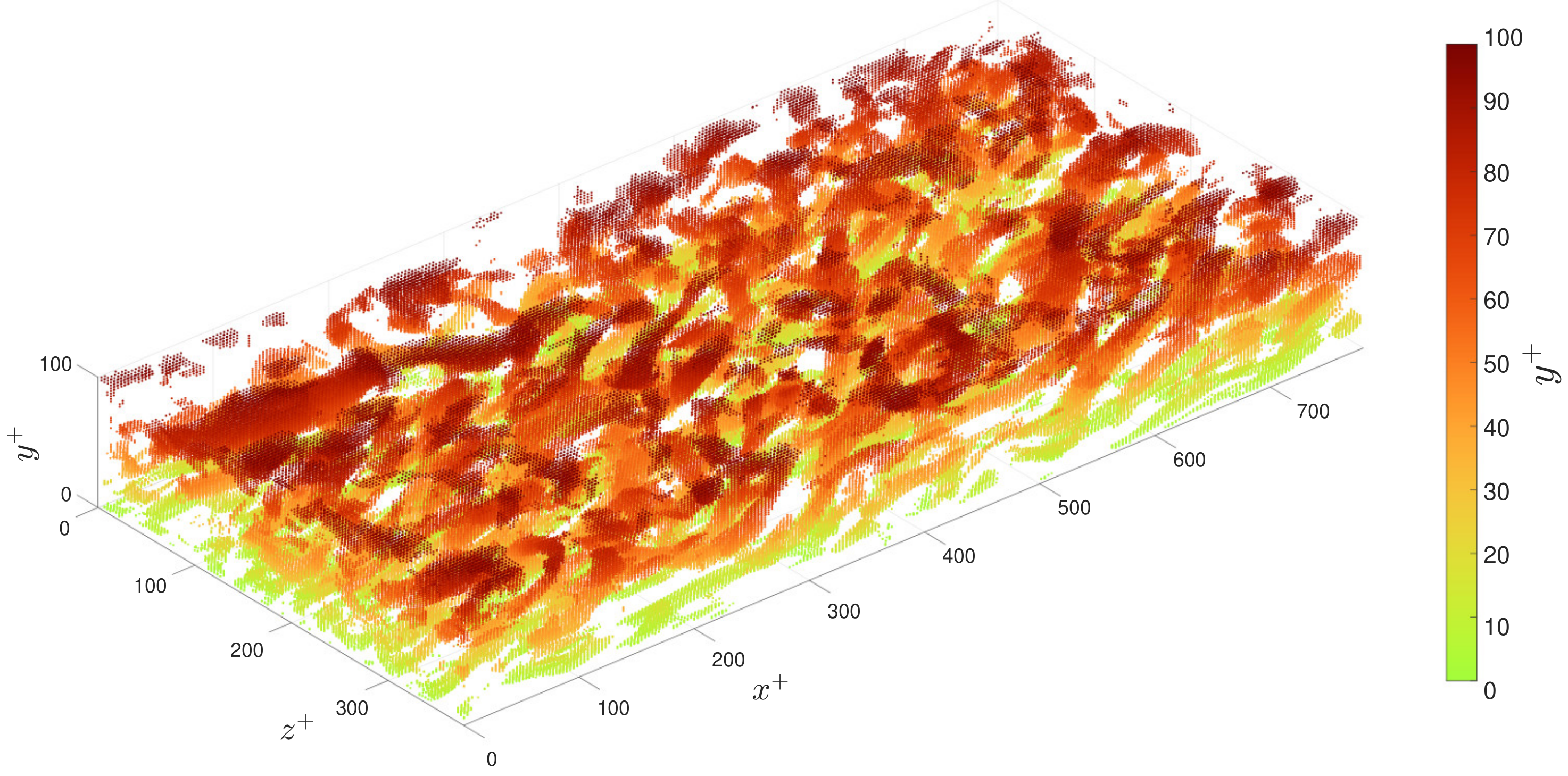}}%
\hfill
\raisebox{-0.5\height}{\includegraphics[width=0.374\textwidth]{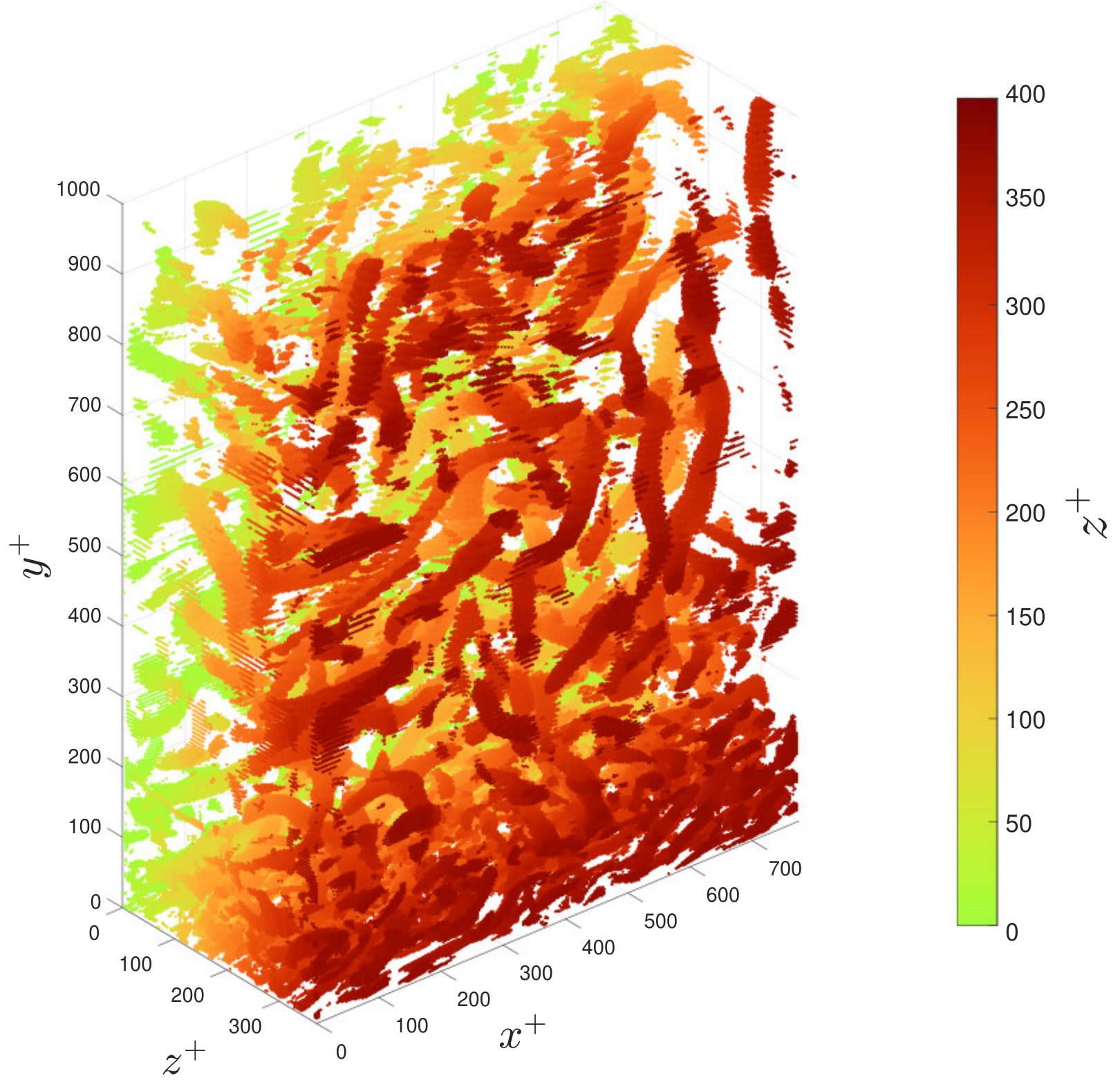}}%
\par\vspace{2pt}
\begin{subfigure}[t]{0.606\textwidth}
  \captionsetup{width=0.616\linewidth}
  \caption{Acceleration-based grid points, \\ $|\boldsymbol{r}|^+<30$ ($0 \leq y^+ \leq 100$)}
\end{subfigure}
\hfill
\begin{subfigure}[t]{0.374\textwidth}
  \caption{Acceleration-based grid points, \\ $|\boldsymbol{r}|^+<30$ ($0 \leq y^+ \leq 1000$)}
\end{subfigure}

\captionsetup{justification=justified}
\caption{Vortex centers in turbulent channel flow obtained using the acceleration-based method (a,b) and the vortex-strength-based method (c,d), and grid points with estimated radii smaller than \(30\) wall units plotted at their original grid locations using the acceleration-based method (e,f).
The near-wall region \(0\leq y^+\leq100\) is shown in the left column (a,c,e) and the region from the wall to the channel center \(0\leq y^+\leq1000\) in the right column (b,d,f).
The vortex centers in (a--d) are colored by the vortex strength \(Q_{\mathrm{sp}}\).
To visualize the structures in (e,f) more clearly, the grid points are colored by \(y^+\) and \(z^+\), respectively.}
\label{fig:turb2}
\end{figure}

The two vortex-center extraction methods are compared in figure~\ref{fig:turb2}.
The acceleration-based and vortex-strength-based results for the near-wall region, \(0<y^+<100\), are shown in figure~\ref{fig:turb2}(a) and (c), respectively, while the corresponding results from the wall to the channel center are shown in (b) and (d).
In figures~\ref{fig:turb2}(a) and (c), the estimated-radius threshold used for data selection is set to \(3\) wall units, whereas in (b) and (d), it is set to \(8\) wall units to account for the larger wall-normal grid spacing away from the wall.
To examine the coherence of the structures surrounding the centerlines, the estimated-radius threshold is increased to \(30\) wall units in figures~\ref{fig:turb2}(e) and (f), and the selected points are plotted at their grid locations rather than at their estimated center locations.
Note that the estimated radius does not necessarily represent the physical distance to the vortex center because the assumption of a constant AGT becomes less valid with increasing distance from the center.

The two extraction methods share several common features.
As in the transitional-flow case, most of the identified vortex centers collectively form filamentary structures.
As shown in figures~\ref{fig:turb2}(e) and (f), increasing the estimated-radius threshold reveals broader tubular structures surrounding these centerlines.
Although these broader tubular structures are shown only for the acceleration-based results, they are also observed in the vortex-strength-based results.
These results indicate that the filamentary centerlines are embedded within coherent tubular regions.
Additionally, in the near-wall region, the centerlines are generally aligned in the streamwise direction, as shown in figures~\ref{fig:turb2}(d) and (e).

The same difference between the two methods is observed here as in the transitional-flow case: most vortex centerlines identified by the acceleration-based method correspond to vortex centerlines identified by the vortex-strength-based method, and the vortex centerlines common to both methods exhibit relatively high vortex strength within their respective neighborhoods.

As a separate test, pressure minima were selected as the centers instead of the zero-acceleration points used in the acceleration-based method.
This test was motivated by the original vortex-core concept associated with pressure minima \citep{hunt1988eddies,jeong1995identification} and was conducted to examine its differences from the acceleration-based method.
Although the pressure-minimum-based method produced clearer centerlines, no substantial differences in the vortex structures or their locations were observed in either the transitional- or turbulent-flow example; for this reason, these results are not shown for comparison.

\subsection{Discussion about acceleration-based and vortex-strength-based methods}
These examples clarify the differences and common features of the acceleration-based and vortex-strength-based methods.
One common feature is that most of the vortex centers identified by either method align to form centerlines, consistent with filamentary vortex structures.
Another common feature is that, in some canonical flows, such as the Euler flow considered here, and for isolated symmetric vortices, such as the Burgers vortex, the location of maximum vortex strength generally coincides with the acceleration-based vortex center or the intuitive center of rotation.
However, this coincidence is not universal: it fails in complex flows and even in simple cases involving rigid-body rotation or Stokes flow.

\subsubsection{Interpretation of vortex-strength-based vortex center
}
The definition of the vortex-strength-based method itself provides a basis for interpreting the identified centers as representing vortices.
This is because a local concentration of vortex strength can indicate the presence of a distinct vortex at that location.
Locally strong vortices are also relevant to the Biot--Savart law, as \citet{chakraborty2005relationships} noted that representing vortices as filaments enables a simple understanding of a substantial portion of the entire flow through this law.
Therefore, regions of locally high vortex strength remain important regardless of whether they can be interpreted as physically distinct vortices.
Nevertheless, applying the Biot--Savart law in practice requires caution because \(Q_{\mathrm{sp}}\) is not vorticity and the identified vortices are generally not infinitesimally thin vortex filaments.
Additionally, vortex strength is derived from relative fluid motion rather than absolute acceleration and is therefore invariant under accelerating reference frame.

\subsubsection{Interpretation of acceleration-based vortex center}

Several examples are examined to clarify the physical interpretation of the acceleration-based vortex center.
In the Stokes-flow example and the rigid-body-rotation case, the acceleration-based method identifies the intuitive center of rotation.
The Stokes-flow example also suggests that interacting vortical motions may produce a combined vortex center whose position depends on their relative strengths.
In the turbulent and transitional flows, the acceleration-based method identifies mainly vortices that appear relatively strong in the vortex-strength-based results, consistent with the behavior observed in the Stokes-flow example.
Additionally, the acceleration-based method naturally incorporates non-local effects by considering acceleration itself, which contains non-local contributions.
Consequently, the identified vortex centers can reflect the influence of the surrounding flow.

Taken together, these observations suggest that an acceleration-based vortex center can be interpreted as the center of a resultant vortex that arises from the combined influence of surrounding vortices, as observed from the reference frame associated with that domain.
Although the ``combined influence of the surrounding flow'' is a more precise description, it is referred to here as the ``combined influence of surrounding vortices'' because this interpretation is consistent with the Stokes-, transitional-, and turbulent-flow examples considered, and because flow fluctuations are generally closely associated with vortical motions, particularly in turbulent flows.
Because the influence of weaker vortices can be obscured by neighboring vortices, the acceleration-based method primarily identifies vortices that are relatively strong within their surrounding vortices, as observed in the transitional and turbulent flows.
The acceleration-based method considers the acceleration that directly governs the Lagrangian motion of fluid elements, whereas the vortex-strength-based method considers only the contribution \(\mathsfbi{A}^{2}\) to the acceleration gradient.
This distinction can explain why, in steady examples such as rigid-body rotation and the Stokes flow considered here, the intuitive center of rotation coincides with the acceleration-based vortex center: the center remains stationary and has zero acceleration in the appropriate reference frame.
In an unsteady flow, this correspondence may no longer hold, and a well-defined center of rotation may not even exist.
Nevertheless, because the acceleration-based method considers acceleration itself, whereas the vortex-strength-based method considers only the \(\mathsfbi{A}^{2}\) contribution to the acceleration gradient, the former may remain more closely aligned with our physical intuition of rotational motion than the latter.

The acceleration-based method is tied to a specific accelerating reference frame, and this dependence may appear problematic; however, it also provides a natural means of introducing a length scale.
For example, let the domain of interest be defined as a sphere of radius \(l\), and let the acceleration of the reference frame be set to the spatially averaged acceleration over this domain.
The resulting acceleration-based vortex centers can then be interpreted as the centers of resultant vortices arising from the combined influence of vortical motions with characteristic scales comparable to or smaller than the domain diameter \(2l\).
Vortical motions with characteristic scales much larger than \(2l\) are expected to produce an approximately uniform acceleration within the domain, and this contribution is largely removed by subtracting the spatially averaged acceleration when defining the reference frame.

Compared with vortex-strength isosurfaces, the acceleration-based method does not require an arbitrarily chosen global threshold.
The acceleration-based method can also identify vortices with low absolute strength if they are sufficiently strong relative to their surrounding vortices and may therefore appropriate to reveal dynamically relevant vortex structures throughout inhomogeneous flows, such as boundary-layer and channel flows.

\subsubsection{Summary}
The physical distinction between the vortex-strength-based and acceleration-based methods can be summarized as that between constituent vortices identified from local concentrations of vortex strength and a resultant vortex observed through acceleration within a specified domain.
Because the acceleration-based method identifies the resultant vortex relative to a reference frame associated with the domain of interest, the selected domain naturally determines the largest characteristic length scale of the vortical motions that influence the resultant vortex.

From a practical perspective, the acceleration-based method may be particularly useful because it directly considers acceleration, which governs the Lagrangian motion of fluid elements, and generally visualizes representative vortices---those with high vortex strength relative to that of surrounding vortices.

\section{Conclusion}

The present study begins by raising two questions about conventional vortex visualization based on criteria such as \(Q\), \(\lambda_2\), and \(\lambda_{ci}\).
The first arises from the existence of multiple vortex-identification criteria, which reflects the absence of a standard measure of vortex strength.
The second is whether a vortex core interpreted as a region of high vortex strength, as commonly visualized using \(Q\), \(\lambda_2\), or \(\lambda_{ci}\), differs from a vortex core interpreted as a pressure-minimum region, as originally conceived.

The first question, arising from the absence of a standard measure of vortex strength, is addressed by introducing the unified criterion \(Q_{\mathrm{sp}}\).
The limitations of the three widely used criteria, \(\lambda_{ci}\), \(Q\), and \(\lambda_2\), are first examined, after which the unified criterion \(Q_{\mathrm{sp}}=\lambda_{ci}^{2}-\lambda_{cr}^{2}\) is derived by combining their key characteristics.
Specifically, \(Q_{\mathrm{sp}}\) adopts the swirling plane identified using \(\lambda_{ci}\), evaluates \(Q\) within that plane to quantify vortex strength, and incorporates the centripetal-acceleration concept underlying the \(\lambda_2\) criterion.
Physically, \(Q_{\mathrm{sp}}\) measures the effective attraction or repulsion between neighboring fluid elements within the swirling plane after excluding the contribution of the material-derivative term of the velocity-gradient tensor.
The unified criterion ensures that the identified centripetal-acceleration pattern is accompanied by swirling motion, which is not guaranteed by \(\lambda_2\).
It also quantifies vortex strength through angular averaging rather than the temporal averaging associated with \(\lambda_{ci}\), and evaluates \(Q\) only within the swirling plane to exclude inappropriate contributions from velocity-gradient components outside that plane.
Finally, the non-orthogonal eigenstructure of the VGT is fully utilized by defining the swirling plane using the complex eigenvectors and subsequently using the remaining real eigenvector as the axis for vortex-center extraction.

To address the second question, concerning the distinction between vortex cores defined as pressure-minimum regions and those defined as regions of high vortex strength, vortex centers are constructed according to the respective definitions.
Conventional vortex-strength isosurfaces are not employed because they do not necessarily provide consistent representations of vortex structures.
The pressure-minimum concept is generalized to a zero-acceleration point at the center of a centripetal-acceleration pattern, allowing it to be applied to a broader range of flows.
Regions of high vortex strength are quantified using the unified criterion \(Q_{\mathrm{sp}}\).
In a three-dimensional flow, the vortex centers, identified as either the centers of centripetal-acceleration patterns or local maxima of \(Q_{\mathrm{sp}}\) within specified planes, generally align to form centerlines, revealing the filamentary structures of the corresponding vortices.
In the Euler-flow example, the proposed method is compared with various alternative vortex-center methods employing different vortex axes and quantities for defining vortex centers.
The limitations observed in these alternative methods further support the proposed unified criterion \(Q_{\mathrm{sp}}\) and the suitability of the selected axis for vortex-center extraction.

The results from the example flows show that the vortex centerlines visualized using the two methods differ considerably.
In the Stokes-flow example, the acceleration-based method identifies the intuitive center of rotation; when global rigid-body rotation is added, it identifies a combined vortex center located between the original vortex and the axis of the added rotation, with its position depending on their relative strengths.
In the transitional- and turbulent-flow examples, the acceleration-based method primarily identifies representative vortices---those exhibiting high vortex strength relative to that of surrounding vortices in the vortex-strength-based visualization---and therefore yields fewer but potentially more significant vortex centerlines.

The physical distinction between the vortex-strength-based and acceleration-based methods can be summarized as that between constituent vortices identified from local concentrations of vortex strength and a resultant vortex identified through acceleration within a specified domain.
The vortex-strength-based method identifies constituent vortices solely from the local VGT and its square, \(\mathsfbi{A}^{2}\).
An acceleration-based vortex center, by contrast, can be interpreted as the center of a resultant vortex arising from the combined influence of the surrounding vortices.
Because the acceleration-based center reflects this combined influence, the method primarily identifies representative vortices---those whose vortex strength is sufficiently high relative to that of the surrounding vortices for their influence not to be obscured.
Because the resultant vortex is identified relative to a reference frame associated with the domain of interest, the selection of that domain naturally determines the largest characteristic length scale considered in identifying the vortex center.
From a practical perspective, the acceleration-based method, rooted in the original vortex-core concept, appears particularly relevant because it directly considers acceleration, and primarily identifies representative vortices.


\section*{Acknowledgement}{
We thank Dr. Sanghyun Ha and Dr. Junshin Park for their kind guidance on performing turbulent flow simulations using the GPU code.
}

\section*{Funding}{
This work was supported by the National Research Foundation of Korea (NRF) under grant
number RS-2023-00282764.
}

\section*{Declaration of interests}{
The authors declare that they have no known competing financial interests or personal
relationships that could have appeared to influence the work reported in this paper.
}

\section*{Data availability statement}
The DNS data of the transitional and turbulent channel flows that support the findings of this study are available from the corresponding author upon reasonable request.

\section*{Author contributions}{
Seungchan Kim: conceptualization, methodology, software, formal analysis, investigation, validation, visualization, data curation, and writing---original draft.
Donghyun You: conceptualization, methodology, writing—original draft, writing—review and editing, supervision, resources, and funding acquisition.
Tamer A. Zaki: conceptualization, methodology, writing—original draft, writing—review and editing, supervision, and validation.
All authors reviewed and approved the final manuscript.
}

\begin{appen}

\section{The Liutex criterion}
\label{app:liutex}
\citet{liu2018rortex} proposed Liutex, originally termed Rortex, which defines the local vortex-axis direction using the real eigenvector of the VGT and quantifies the rotational strength as the irremovable rotational component of vorticity.
Equivalently, the Liutex magnitude is twice the minimum angular velocity in the plane perpendicular to this axis.
This plane generally differs from the two-dimensional swirling plane spanned by the real and imaginary parts of a complex eigenvector, whose normal is the real left eigenvector of the VGT.
By choosing a coordinate system whose third basis vector is aligned with the real eigenvector of the VGT, the VGT can be written as

\begin{equation}
\mathsfbi{A}
=
\left[A_{ij}\right]
=
\left[\frac{\partial u_i}{\partial x_j}\right]
=
\begin{pmatrix}
a' & b' & 0 \\
c' & d' & 0 \\
e'' & f'' & g
\end{pmatrix}.
\end{equation}

As in the $\lambda_{ci}$ case, the corresponding quantities are defined as
\begin{equation}
\alpha'=\frac{a'+d'}{2},\qquad
\beta'=\frac{b'-c'}{2},\qquad
\gamma'=\frac{1}{2}\sqrt{(a'-d')^2+(b'+c')^2}.
\end{equation}

Based on these quantities, the Liutex magnitude can be written as $R=2(|\beta'|-\gamma')$.
Although the tensor components change from $a$, $b$, $c$, and $d$, defined in the swirling plane, to $a'$, $b'$, $c'$, and $d'$, the relations $\alpha'=\alpha$ and ${\beta'}^2-{\gamma'}^2=\beta^2-\gamma^2=\lambda_{ci}^2$ remain invariant.
This invariance follows because $\alpha$ and $\beta^2-\gamma^2$ correspond to the real part and the squared imaginary part of the complex eigenvalues, respectively, and the left and right eigenvalues are identical.
Therefore, when a zero threshold is used, $\lambda_{ci}$ and Liutex classify a region identically as either vortical or non-vortical, although their magnitudes differ.

\section{Calculation of the angular average of the angular velocity}
\label{app:average_angular_vorticity}
Again, by choosing a coordinate system whose third basis vector is aligned with the real left eigenvector of the VGT, the VGT can be written as in equation~\eqref{eq:VGT_left}, with \(\alpha\), \(\beta\), and \(\gamma\) defined in equation~\eqref{eq:alphabetagamma}.
In the swirling plane spanned by the first two basis vectors, \(x_1\) and \(x_2\), the position vector \(\boldsymbol{r}\) can be expressed as \(\boldsymbol{r}=(|\boldsymbol{r}|\cos\theta,|\boldsymbol{r}|\sin\theta)\), where \(\theta\) denotes the angle between the \(x_1\)-direction and \(\boldsymbol{r}\).
The angular velocity within the swirling plane can then be written as
\begin{align}
\dot{\theta}
&=\frac{\boldsymbol{r}\times\boldsymbol{u}}{|\boldsymbol{r}|^2}\\
&=\frac{x_1u_2-x_2u_1}{|\boldsymbol{r}|^2}\\
&=\frac{x_1Ax_2-x_2Ax_1}{|\boldsymbol{r}|^2}\\
&=\cos\theta\left(c\cos\theta+d\sin\theta\right) -\sin\theta\left(a\cos\theta+b\sin\theta\right)\\
&=-\beta+\frac{b+c}{2}\cos(2\theta)-\frac{a-d}{2}\sin(2\theta).
\end{align}
Let $\frac{b+c}{2}=\gamma\cos(2\theta_0)$ and $\frac{a-d}{2}=\gamma\sin(2\theta_0)$.
Then, $\dot{\theta}(\theta)$ can be written as
\begin{equation}
\dot{\theta}(\theta)=-\beta+\gamma\cos[2(\theta+\theta_0)].
\end{equation}
Then, the angular average of the angular velocity is given by
\begin{align}
\overline{\dot{\theta}}&=\frac{1}{2\pi}\int_0^{2\pi}\dot{\theta}(\theta)\,d\theta\\
&=\frac{1}{2\pi}\int_0^{2\pi}\left\{-\beta+\gamma\cos[2(\theta+\theta_0)]\right\}\,d\theta\\
&=-\beta.
\end{align}
For a complex-conjugate pair of eigenvalues to exist, the following condition must be satisfied:
\begin{equation}
|\beta|>\gamma.
\end{equation}
In this case, the minimum and maximum magnitudes of the angular velocity are $|\beta|-\gamma$ and $|\beta|+\gamma$, respectively.
The minimum magnitude of the angular velocity corresponds to the Liutex defined by \citet{liu2018rortex}, although the two are not exactly equivalent because Liutex is defined in the plane normal to the real right eigenvector rather than in the 2D swirling plane normal to the real left eigenvector.

\section{Calculation of the angular average of the radial acceleration gradient.}
\label{app:average_angular_acceleration}
As in the calculation of the average angular velocity in Appendix~\ref{app:average_angular_vorticity}, the notation introduced in Eqs.~\eqref{eq:alphabetagamma} and \eqref{eq:lci} is used.
In the swirling plane spanned by the first two basis vectors, \(x_1\) and \(x_2\), the position vector \(\boldsymbol{r}\) can be expressed as \(\boldsymbol{r}=(|\boldsymbol{r}|\cos\theta,|\boldsymbol{r}|\sin\theta)\), where \(\theta\) denotes the angle between the \(x_1\)-direction and \(\boldsymbol{r}\).
The relative acceleration between the origin and the position \(\boldsymbol{r}\) within the swirling plane, induced by the \(\mathsfbi{A}^2\) term, is given by \(\boldsymbol{a}_{\mathrm{rel}}=\mathsfbi{A}^2\boldsymbol{r}\).
The radial acceleration gradient, evaluated as the radial component of the acceleration difference per unit distance from the origin, is then given by
\begin{equation}
a_r = \frac{\boldsymbol{a}_{\mathrm{rel}}\boldsymbol{\cdot}\boldsymbol{r}}{|\boldsymbol{r}|^2} 
= \boldsymbol{e}_r^{\mathsf T}\mathsfbi{A}^2\boldsymbol{e}_r, \qquad
\boldsymbol{e}_r = (\cos\theta,\sin\theta).
\end{equation}
It can be written as
\begin{equation}
a_r(\theta) = (a^2+bc)\cos^2\theta + (b+c)(a+d)\cos\theta\sin\theta + (d^2+bc)\sin^2\theta.
\end{equation}
Let \((b+c)/2=\gamma\cos(2\theta_0)\) and \((a-d)/2=\gamma\sin(2\theta_0)\). Then,
\begin{equation}
a=\alpha+\gamma\sin(2\theta_0),\;
d=\alpha-\gamma\sin(2\theta_0),\;
b=\beta+\gamma\cos(2\theta_0),\;
c=-\beta+\gamma\cos(2\theta_0).
\end{equation}
Substituting these expressions gives
\begin{align}
a_r(\theta)
={}&
\left(
\alpha^2+\gamma^2-\beta^2
+
2\alpha\gamma\sin(2\theta_0)
\right)\cos^2\theta
\nonumber\\
&{}+
4\alpha\gamma\cos(2\theta_0)
\cos\theta\sin\theta
\nonumber\\
&{}+
\left(
\alpha^2+\gamma^2-\beta^2
-
2\alpha\gamma\sin(2\theta_0)
\right)\sin^2\theta \\
={}& 
\alpha^2+\gamma^2-\beta^2
+
2\alpha\gamma\sin\left[2(\theta+\theta_0)\right].
\end{align}
Taking the angular average of the radial acceleration gradient gives
\begin{align}
\overline{a}_r
&=
\frac{1}{2\pi}
\int_0^{2\pi}
a_r(\theta)\,d\theta\\
&=
\frac{1}{2\pi}
\int_0^{2\pi}
\left\{
\alpha^2+\gamma^2-\beta^2
+
2\alpha\gamma\sin\left[2(\theta+\theta_0)\right]
\right\}
\,d\theta\\
&=
\alpha^2+\gamma^2-\beta^2.
\end{align}
\end{appen}

\bibliographystyle{preprint}
\bibliography{references}

\begin{thebibliography}{32}
\expandafter\ifx\csname natexlab\endcsname\relax\def\natexlab#1{#1}\fi
\def\au#1{#1} \def\ed#1{#1} \def\yr#1{#1}\def\at#1{#1}\def\jt#1{\textit{#1}}
  \def\bt#1{#1}\def\bvol#1{\textbf{#1}} \def\vol#1{#1} \def\pg#1{#1}
  \def\publ#1{#1}\def\arxiv#1{#1}\def\org#1{#1}\def\st#1{\textit{#1}}

\bibitem[Adrian(2007)]{adrian2007hairpin}
{\sc \au{Adrian, Ronald~J.}} \yr{2007}  \at{Hairpin vortex organization in wall
  turbulence}.  \jt{Physics of Fluids}  \bvol{19}~(4),  \pg{041301}.

\bibitem[Adrian {\em et~al.\/}(2000)Adrian, Meinhart \&
  Tomkins]{adrian2000vortex}
{\sc \au{Adrian, Ronald~J.}, \au{Meinhart, Carl~D.} \& \au{Tomkins,
  Christopher~D.}} \yr{2000}  \at{Vortex organization in the outer region of
  the turbulent boundary layer}.  \jt{Journal of Fluid Mechanics}  \bvol{422},
  \pg{1--54}.

\bibitem[Biringen \& Danabasoglu(1988)]{biringen1988solution}
{\sc \au{Biringen, Sedat} \& \au{Danabasoglu, Gokhan}} \yr{1988}  \bt{Solution
  of the {O}rr--{S}ommerfeld equation for the {B}lausius boundary-layer
  documentation of program {ORRBL} and a test case}. {NASA} contractor report.
  \org{NASA Langley Research Center}, Hampton, VA.

\bibitem[Cantwell(1992)]{cantwell1992exact}
{\sc \au{Cantwell, Brian~J.}} \yr{1992}  \at{Exact solution of a restricted
  {E}uler equation for the velocity gradient tensor}.  \jt{Physics of Fluids A:
  Fluid Dynamics}  \bvol{4}~(4),  \pg{782--793}.

\bibitem[Canuto {\em et~al.\/}(2006)Canuto, Hussaini, Quarteroni \&
  Zang]{canuto2006spectral}
{\sc \au{Canuto, Claudio}, \au{Hussaini, M.~Yousuff}, \au{Quarteroni, Alfio} \&
  \au{Zang, Thomas~A.}} \yr{2006} {\em Spectral Methods: Fundamentals in Single
  Domains\/}.  \publ{Springer}.

\bibitem[Chakraborty {\em et~al.\/}(2005)Chakraborty, Balachandar \&
  Adrian]{chakraborty2005relationships}
{\sc \au{Chakraborty, Pinaki}, \au{Balachandar, S.} \& \au{Adrian, Ronald~J.}}
  \yr{2005}  \at{On the relationships between local vortex identification
  schemes}.  \jt{Journal of Fluid Mechanics}  \bvol{535},  \pg{189--214}.

\bibitem[Chevillard {\em et~al.\/}(2008)Chevillard, Meneveau, Biferale \&
  Toschi]{chevillard2008modeling}
{\sc \au{Chevillard, L.}, \au{Meneveau, C.}, \au{Biferale, L.} \& \au{Toschi,
  F.}} \yr{2008}  \at{Modeling the pressure {H}essian and viscous {L}aplacian
  in turbulence: comparisons with direct numerical simulation and implications
  on velocity gradient dynamics}.  \jt{Physics of Fluids}  \bvol{20}~(10),
  \pg{101504}.

\bibitem[Chong {\em et~al.\/}(1990)Chong, Perry \& Cantwell]{chong1990general}
{\sc \au{Chong, Min~S.}, \au{Perry, Anthony~E.} \& \au{Cantwell, Brian~J.}}
  \yr{1990}  \at{A general classification of three-dimensional flow fields}.
  \jt{Physics of Fluids A: Fluid Dynamics}  \bvol{2}~(5),  \pg{765--777}.

\bibitem[Dombre {\em et~al.\/}(1986)Dombre, Frisch, Greene, H{\'e}non, Mehr \&
  Soward]{dombre1986chaotic}
{\sc \au{Dombre, Thierry}, \au{Frisch, Uriel}, \au{Greene, John~M.},
  \au{H{\'e}non, Michel}, \au{Mehr, A.} \& \au{Soward, Andrew~M.}} \yr{1986}
  \at{Chaotic streamlines in the {ABC} flows}.  \jt{Journal of Fluid Mechanics}
   \bvol{167},  \pg{353--391}.

\bibitem[Ha {\em et~al.\/}(2021)Ha, Park \& You]{ha_multi-gpu_2021}
{\sc \au{Ha, Sanghyun}, \au{Park, Junshin} \& \au{You, Donghyun}} \yr{2021}
  \at{A multi-{GPU} method for {ADI}-based fractional-step integration of
  incompressible {N}avier--{S}tokes equations}.  \jt{Computer Physics
  Communications}  \bvol{265},  \pg{107999}.

\bibitem[Haller(2005)]{haller2005objective}
{\sc \au{Haller, George}} \yr{2005}  \at{An objective definition of a vortex}.
  \jt{Journal of Fluid Mechanics}  \bvol{525},  \pg{1--26}.

\bibitem[Haller {\em et~al.\/}(2016)Haller, Hadjighasem, Farazmand \&
  Huhn]{haller2016defining}
{\sc \au{Haller, George}, \au{Hadjighasem, Alireza}, \au{Farazmand, Mohammad}
  \& \au{Huhn, Florian}} \yr{2016}  \at{Defining coherent vortices objectively
  from the vorticity}.  \jt{Journal of Fluid Mechanics}  \bvol{795},
  \pg{136--173}.

\bibitem[Hunt {\em et~al.\/}(1988)Hunt, Wray \& Moin]{hunt1988eddies}
{\sc \au{Hunt, Julian C.~R.}, \au{Wray, Alan~A.} \& \au{Moin, Parviz}}
  \yr{1988} Eddies, streams, and convergence zones in turbulent flows.  \bt{In
  {\em Studying Turbulence Using Numerical Simulation Databases, 2. Proceedings
  of the 1988 Summer Program\/}},  \pg{pp. 193--208}. Center for Turbulence
  Research, report CTR-S88.

\bibitem[Jeong \& Hussain(1995)]{jeong1995identification}
{\sc \au{Jeong, Jinhee} \& \au{Hussain, Fazle}} \yr{1995}  \at{On the
  identification of a vortex}.  \jt{Journal of Fluid Mechanics}  \bvol{285},
  \pg{69--94}.

\bibitem[Jeong {\em et~al.\/}(1997)Jeong, Hussain, Schoppa \&
  Kim]{jeong1997coherent}
{\sc \au{Jeong, Jinhee}, \au{Hussain, Fazle}, \au{Schoppa, Wade} \& \au{Kim,
  John}} \yr{1997}  \at{Coherent structures near the wall in a turbulent
  channel flow}.  \jt{Journal of Fluid Mechanics}  \bvol{332},  \pg{185--214}.

\bibitem[Jim{\'e}nez {\em et~al.\/}(1993)Jim{\'e}nez, Wray, Saffman \&
  Rogallo]{jimenez1993structure}
{\sc \au{Jim{\'e}nez, Javier}, \au{Wray, Alan~A.}, \au{Saffman, Philip~G.} \&
  \au{Rogallo, Robert~S.}} \yr{1993}  \at{The structure of intense vorticity in
  isotropic turbulence}.  \jt{Journal of Fluid Mechanics}  \bvol{255},
  \pg{65--90}.

\bibitem[Kida \& Miura(1998)]{kida1998identification}
{\sc \au{Kida, Shigeo} \& \au{Miura, Hideaki}} \yr{1998}  \at{Identification
  and analysis of vortical structures}.  \jt{European Journal of Mechanics --
  B/Fluids}  \bvol{17}~(4),  \pg{471--488}.

\bibitem[Kim {\em et~al.\/}(2025)Kim, Kim, Ha \& You]{kim2025highly}
{\sc \au{Kim, Seungchan}, \au{Kim, Jihoo}, \au{Ha, Sanghyun} \& \au{You,
  Donghyun}} \yr{2025} A highly scalable {TDMA} for {GPU}s and its application
  to flow solver optimization. arXiv preprint arXiv:2509.03933.

\bibitem[Lee \& Zaki(2017)]{lee2017simulations}
{\sc \au{Lee, Sang~Jin} \& \au{Zaki, Tamer~A.}} \yr{2017}  \at{Simulations of
  natural transition in viscoelastic channel flow}.  \jt{Journal of Fluid
  Mechanics}  \bvol{820},  \pg{232--262}.

\bibitem[Liu {\em et~al.\/}(2018)Liu, Gao, Tian \& Dong]{liu2018rortex}
{\sc \au{Liu, Chaoqun}, \au{Gao, Yisheng}, \au{Tian, Shuling} \& \au{Dong,
  Xiangrui}} \yr{2018}  \at{Rortex---{A} new vortex vector definition and
  vorticity tensor and vector decompositions}.  \jt{Physics of Fluids}
  \bvol{30}~(3),  \pg{035103}.

\bibitem[Meneveau(2011)]{meneveau2011lagrangian}
{\sc \au{Meneveau, Charles}} \yr{2011}  \at{{L}agrangian dynamics and models of
  the velocity gradient tensor in turbulent flows}.  \jt{Annual Review of Fluid
  Mechanics}  \bvol{43},  \pg{219--245}.

\bibitem[Miura \& Kida(1997)]{miura1997identification}
{\sc \au{Miura, Hideaki} \& \au{Kida, Shigeo}} \yr{1997}  \at{Identification of
  tubular vortices in turbulence}.  \jt{Journal of the Physical Society of
  Japan}  \bvol{66}~(5),  \pg{1331--1334}.

\bibitem[Panton(2024)]{panton2024incompressible}
{\sc \au{Panton, Ronald~L.}} \yr{2024} {\em Incompressible Flow\/}, 5th edn.
  \publ{John Wiley \& Sons}.

\bibitem[She {\em et~al.\/}(1991)She, Jackson \& Orszag]{she1991structure}
{\sc \au{She, Zhen-Su}, \au{Jackson, Eric} \& \au{Orszag, Steven~A.}} \yr{1991}
   \at{Structure and dynamics of homogeneous turbulence: models and
  simulations}.  \jt{Proceedings of the Royal Society of London. Series A:
  Mathematical and Physical Sciences}  \bvol{434}~(1890),  \pg{101--124}.

\bibitem[Stone {\em et~al.\/}(1991)Stone, Nadim \& Strogatz]{stone1991chaotic}
{\sc \au{Stone, Howard~A.}, \au{Nadim, Ali} \& \au{Strogatz, Steven~H.}}
  \yr{1991}  \at{Chaotic streamlines inside drops immersed in steady {S}tokes
  flows}.  \jt{Journal of Fluid Mechanics}  \bvol{232},  \pg{629--646}.

\bibitem[Sujudi \& Haimes(1995)]{sujudi1995identification}
{\sc \au{Sujudi, David} \& \au{Haimes, Robert}} \yr{1995} Identification of
  swirling flow in {3-D} vector fields.  \bt{In {\em 12th AIAA Computational
  Fluid Dynamics Conference\/}}. AIAA Paper 95-1715.

\bibitem[Tanahashi {\em et~al.\/}(2001)Tanahashi, Iwase \&
  Miyauchi]{tanahashi2001appearanceand}
{\sc \au{Tanahashi, M.}, \au{Iwase, S.} \& \au{Miyauchi, T.}} \yr{2001}
  \at{Appearance and alignment with strain rate of coherent fine scale eddies
  in turbulent mixing layer}.  \jt{Journal of Turbulence}  \bvol{2},  \pg{N6}.

\bibitem[Vieillefosse(1982)]{vieillefosse1982local}
{\sc \au{Vieillefosse, P.}} \yr{1982}  \at{Local interaction between vorticity
  and shear in a perfect incompressible fluid}.  \jt{Journal de Physique}
  \bvol{43}~(6),  \pg{837--842}.

\bibitem[Wang {\em et~al.\/}(2022)Wang, Shen, Mang, Gao, Wang \&
  Wang]{wang2022novel}
{\sc \au{Wang, Chengyue}, \au{Shen, Yunian}, \au{Mang, Shanshan}, \au{Gao, Qi},
  \au{Wang, Zhongyi} \& \au{Wang, Jinjun}} \yr{2022}  \at{A novel algorithm for
  visualizing and quantifying vortices in complex {3D} flows based on marching
  and converging vortex atoms}.  \jt{Physics of Fluids}  \bvol{34}~(11),
  \pg{115131}.

\bibitem[Yu {\em et~al.\/}(2026)Yu, Alvarez \& Liu]{yu2026new}
{\sc \au{Yu, Yifei}, \au{Alvarez, Oscar} \& \au{Liu, Chaoqun}} \yr{2026}  \at{A
  new automatic vortex ({Liutex}) core line identification method in
  boundary-layer transition}.  \jt{Physics of Fluids}  \bvol{38}~(7).

\bibitem[Zhou {\em et~al.\/}(1999)Zhou, Adrian, Balachandar \&
  Kendall]{zhou1999mechanisms}
{\sc \au{Zhou, J.}, \au{Adrian, R.~J.}, \au{Balachandar, S.} \& \au{Kendall,
  T.~M.}} \yr{1999}  \at{Mechanisms for generating coherent packets of hairpin
  vortices in channel flow}.  \jt{Journal of Fluid Mechanics}  \bvol{387},
  \pg{353--396}.

\bibitem[Zhu \& Xi(2019)]{zhu2019vortex}
{\sc \au{Zhu, Lu} \& \au{Xi, Li}} \yr{2019}  \at{Vortex axis tracking by
  iterative propagation ({VATIP}): a method for analysing three-dimensional
  turbulent structures}.  \jt{Journal of Fluid Mechanics}  \bvol{866},
  \pg{169--215}.

\end{thebibliography}

\end{document}